%% file: ms.tex
\pdfoutput=1
\documentclass[sigconf, nonacm]{acmart}

\newcommand\vldbdoi{XX.XX/XXX.XX}
\newcommand\vldbpages{XXX-XXX}
\newcommand\vldbvolume{14}
\newcommand\vldbissue{1}
\newcommand\vldbyear{2020}
\newcommand\vldbauthors{\authors}
\newcommand\vldbtitle{\shorttitle} 
\newcommand\vldbavailabilityurl{http://vldb.org/pvldb/format_vol14.html}
\newcommand\vldbpagestyle{plain}

\usepackage{graphicx}
\usepackage[inline]{enumitem}
\usepackage{wrapfig}
\usepackage{hyperref, xcolor}

\usepackage{lmodern}
\usepackage[T1]{fontenc}
\usepackage[utf8]{inputenc}
\usepackage{multirow}
\usepackage{xcolor}
\usepackage{bbding}
\usepackage{pifont}
\usepackage{wasysym}
\usepackage{array,multirow}
\usepackage{cleveref}
\usepackage{soul}
\usepackage{dblfloatfix}

\usepackage{url}
\usepackage{pifont}

\usepackage{listings}
\usepackage{xcolor}

\definecolor{codegreen}{rgb}{0,0.6,0}
\definecolor{codegray}{rgb}{0.5,0.5,0.5}
\definecolor{codepurple}{rgb}{0.58,0,0.82}
\definecolor{backcolour}{rgb}{0.95,0.95,0.92}

\usepackage{array}
\usepackage{booktabs}
\usepackage{colortbl}
\usepackage{multirow}
\usepackage{amsmath}
\usepackage{amsfonts}
\usepackage{caption}
\usepackage{subcaption}
\usepackage{mathtools}
\usepackage{algorithm}
\usepackage{algorithmic}
\usepackage{pbox}
\begin{document}

\title{Scaling Hyperledger Fabric Using Pipelined Execution and Sparse Peers}
\author{Parth Thakkar}
\authornote{Currently employed at Microsoft Research, India}
\affiliation{%
  \institution{IBM Research, India}
  \streetaddress{}
}
\email{thakkar.parth.d@gmail.com}

\author{Senthilnathan Natarajan}
\authornote{Committer, Hyperledger Fabric}
\affiliation{%
  \institution{IBM Research, India}
  \streetaddress{}
}
\email{senthil.nathan.7@gmail.com}

\input{abstract.tex}
\maketitle






\pagestyle{\vldbpagestyle}
\begingroup\small\noindent\raggedright\textbf{PVLDB Reference Format:}\\
\vldbauthors. \vldbtitle. PVLDB, \vldbvolume(\vldbissue): \vldbpages, \vldbyear.\\
\href{https://doi.org/\vldbdoi}{doi:\vldbdoi}
\endgroup
\begingroup
\renewcommand\thefootnote{}\footnote{\noindent
This work is licensed under the Creative Commons BY-NC-ND 4.0 International License. Visit \url{https://creativecommons.org/licenses/by-nc-nd/4.0/} to view a copy of this license. For any use beyond those covered by this license, obtain permission by emailing \href{mailto:info@vldb.org}{info@vldb.org}. Copyright is held by the owner/author(s). Publication rights licensed to the VLDB Endowment. \\
\raggedright Proceedings of the VLDB Endowment, Vol. \vldbvolume, No. \vldbissue\ %
ISSN 2150-8097. \\
\href{https://doi.org/\vldbdoi}{doi:\vldbdoi} \\
}\addtocounter{footnote}{-1}\endgroup

\ifdefempty{\vldbavailabilityurl}{}{
\vspace{.3cm}
}

\input{introduction.tex}

\input{background-and-motivation.tex}
\input{design.tex}

\input{evaluation.tex}
\input{relatedwork.tex}


\sloppy
\bibliographystyle{abbrv}
\bibliography{ms}  

\end{document}

%% file: abstract.tex
\begin{abstract}
    Permissioned blockchains are becoming popular as data management systems in
    the enterprise setting. Compared to traditional distributed databases, blockchain platforms
    provide increased security guarantees but significantly
    lower performance. Further, these platforms are quite expensive to run for the low throughput
    they provide. The following are two ways to improve performance and reduce cost:
    (1) make the system utilize allocated resources efficiently;
    (2) allow rapid and dynamic scaling of allocated resources based on load.
    We explore both of these in this work.

    We first investigate the reasons for the poor performance and scalability of
    the dominant permissioned blockchain flavor called \texttt{Execute-Order-Validate} (\texttt{EOV}).
    We do this by studying the scaling characteristics of Hyperledger Fabric,
    a popular EOV platform, using \textit{vertical scaling} and
    \textit{horizontal scaling}.
    We find that the transaction throughput scales very poorly with these techniques.
    At least in the permissioned setting, the real bottleneck is transaction processing, not the consensus protocol.
    With vertical scaling, the allocated vCPUs go under-utilized.
    %
    In contrast, with horizontal scaling, the allocated resources get wasted due to redundant work
    across nodes within an organization.
    
    \par
    To mitigate the above concerns, we first improve resource efficiency
    by (a) improving CPU utilization with a pipelined execution of validation \& commit phases;
    (b) avoiding redundant work across nodes by introducing a new type of peer
    node called \textit{sparse peer} that selectively commits transactions. We further propose a technique that
    enables the rapid scaling of resources.
    Our implementation -- SmartFabric, built on top of Hyperledger Fabric
    demonstrates 3$\times$ higher throughput, 12-26$\times$ faster scale-up time,
    and provides Fabric's throughput at 50\% to 87\% lower cost.
\end{abstract}

%% file: introduction.tex
\section{Introduction}
Blockchain technologies gained popularity as they provide a way to eliminate
the intermediary and decentralize the application.
A blockchain is a ledger that records transactions and data which {are} replicated across
multiple peer nodes where {nodes do not trust one another}. Each node holds
an identical copy of the ledger as a chain of blocks. Each block is a
logical sequence of transactions and encloses the hash of its immediate
previous block, thereby guaranteeing the ledger's immutability.
A block is created by executing a consensus
protocol among the nodes.

\par
For enterprise use-cases that involve
interaction between multiple organizations, a permissioned blockchain network
is suitable as it supports authenticated participants and privacy.
Different organizations own nodes in a permissioned network,
and transaction logic gets implemented using smart-contracts.
There are two dominant
blockchain transaction models: Execute-Order-Validate (\texttt{EOV}) and Order-Execute
(\texttt{OE}).
Hyperledger Fabric~\cite{hf} and Quorum~\cite{quorum},
which follow the \texttt{EOV} and \texttt{OE} model, respectively, are the two most
popular permissioned blockchain platforms, as mentioned in Forbes Blockchain 50~\cite{forbes}.
This paper focuses on the \texttt{EOV} model.
\par
Hyperledger Fabric's performance is of significant concern for enterprises
due to steady growth in network usage. For example, to track the provenance of ingredients
used in food products, such as protein bars, chocolates, other packaged foods, we need to store lots and
lots of records at a high rate.
In the current form, permissioned blockchain platforms cannot provide the performance
needed by large provenance use-cases and finance industry---stock exchanges, credit card
companies, such as Visa~\cite{visa-annual-report}, and mobile payment platforms,
such as AliPay~\cite{alipay} (a peak of 325k tps). Further, with innovations in
tokens~\cite{token1, token2} and related applications such as decentralized
marketplace~\cite{origami-network, open-bazaar, particl, electrumdark},
the throughput requirement will only increase. Hence, it is necessary to improve
the performance of Fabric to support ongoing growth proactively. Even with the current
performance provided by Fabric, nodes get overprovisioned to satisfy a peak load as
there exists no mechanism to scale up/down a network. As a result, operational cost
increases unnecessarily.
\par
Although many efforts~\cite{perf-ssi, perf-mascots, perf-noblock, perf-priority,
		perf-sharding, perf-concurrency, xox, fastfabric, tps}
have proposed various optimizations to improve Fabric's performance, none have comprehensively
studied the impact of scaling techniques, such as \textit{vertical} and
\textit{horizontal} scaling, to identify bottlenecks.
Hence, in this paper, we study the efficiency of various scaling
techniques and identify bottlenecks. Then, we re-architect Fabric transparently
to improve performance.
In general, the consensus layer is assumed to be the bottleneck.
Though that is valid for the permissionless blockchains such as Ethereum~\cite{ethereum},
Bitcoin~\cite{bitcoin} due to Proof of Work (PoW) consensus, we found that the validation and
commit of a block is the bottleneck in Fabric, not the consensus layer as it uses Raft~\cite{raft}.
Our four major contributions are:
\begin{enumerate}[topsep=0pt,itemsep=-0pt,leftmargin=15pt]
		\item We conducted experiments to understand the scalability of Fabric using vertical and horizontal scaling.
			  We find that the transaction throughput scales very poorly with these techniques.
			  With vertical scaling, the allocated vCPUs go under-utilized, while with horizontal scaling, the allocated
              resources get wasted due to redundant work across nodes within an organization.
		\item We re-architected Fabric to enable pipelined execution of validation \& commit phases without violating the serializability
				isolation. We also facilitated
				the validation phase to validate multiple transactions, which belong to different blocks, in parallel. We achieve this by introducing a
				\textit{waiting-transactions dependency graph} that tracks 7 types of dependencies between transactions.
				As a result, the performance improved by 1.4$\times$ while increasing the CPU utilization from 40\% to 70\%.
		\item We introduced a new type of
				node called \textit{sparse peer} that selectively commits transactions to avoid the duplication of CPU
                \& IO intensive {tasks}. Thus, the performance improved by 2.4$\times$. With both \textit{sparse peer}
                and \textit{pipelined execution of phases}, the performance improved by
				3$\times$.
		\item We built an auto-scaling framework that can split a full peer into multiple \textit{sparse peers} or merge multiple
				\textit{sparse peers} into a full peer. This helps in scaling up
				a network quickly to handle an overload situation and reduce transaction invalidation. Our approach reduced
			    the scale-up time of a network by 12-26$\times$ while increasing the scale-down time slightly.
\end{enumerate}
The remainder of this paper is structured as follows. \S\ref{background-and-motivation} provides background on Hyperledger
Fabric and motivates our work by performing various experiments. 
\S\ref{parallel-validation} and \S\ref{sparse-peer} describe our proposed architecture for Fabric.
\S\ref{evaluation} evaluates our proposed architecture to showcase the
improvement achieved and \S\ref{relatedwork} presents related work. 

%% file: background-and-motivation.tex
\vspace{-.4cm}
\section{Background and Motivation}\label{background-and-motivation}
\subsection{Hyperledger Fabric}\label{hf}
\begin{figure}
    \begin{center}
			\includegraphics[scale=0.35]{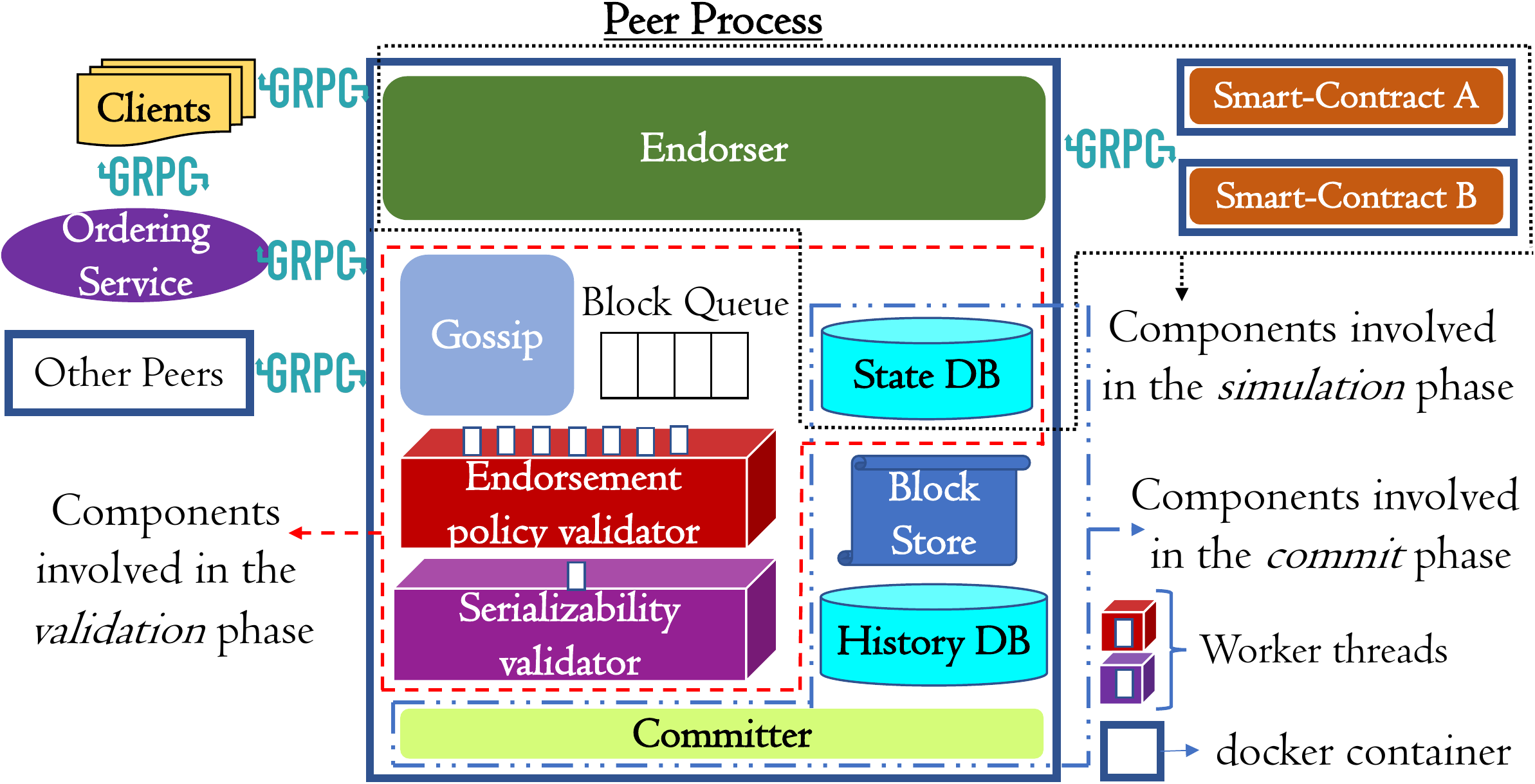}
  \vspace{-0.7cm}
		\caption{\small{Various components in a peer}}
		\label{fig:peer}
 \end{center}
  \vspace{-0.85cm}
\end{figure}

Hyperledger Fabric consists of three entities---client,
peer, and orderer. The transaction flow in Fabric
involves all three entities and comprises four phases---
execution, ordering, validation, and commit, {as depicted
in~\cite{hf}}. Figure~\ref{fig:peer} shows various peer's components
that involve in the execution of all these phases except the
ordering phase. The
communication between different entities happens via
gRPC~\cite{grpc}.
\par
\textbf{Phase 1---Simulation.} {A client submits a transaction proposal to a peer to invoke a
smart-contract, which implements the transaction logic. 
The \textit{smart-contract} executes the logic and reads/writes the states it
manages by issuing \texttt{get()}, \texttt{put()}, and \texttt{range()} commands. 
Once the transaction execution completes,
the \textit{endorser} cryptographically signs/endorses the transaction response
(which includes the transaction proposal, read-write set, and range query info)
before sending it to the client. The read-write set includes all the keys read and written
by the contract. The range query info includes the start and end key passed to the
range query.}
The client can submit the
transaction proposal to multiple peers simultaneously depending upon the
endorsement policy~\cite{endorsement1}
defined for the \textit{smart-contract}.
Each \textit{smart-contract} maintains its blockchain states. One \textit{smart-contract}
can invoke other contracts to read/write their blockchain states.

\textbf{Phase 2---Ordering.}
The client submits the endorsed transaction response to the ordering
service. The ordering service, which consists of orderer nodes from different
organizations, employs a consensus protocol~\cite{raft} to order the received
transactions and create a block. Each block has a sequence number called
\textit{block number}, the hash of the previous block, the hash of the current block,
a list of ordered transactions (i.e., \textit{commit order}), and the orderer's
signature. The orderer nodes broadcast these blocks to peers.
\par
\textbf{Phase 3---Validation.} The \textit{gossip} component of the peer
receives blocks and stores them in the \textit{block queue}. Before committing a block,
the peer runs every transaction in the block through the following two validators:
(1) The \texttt{endorsement policy validator} verifies
that the transaction response has been signed by enough organizations as specified in the contract's endorsement policy.
If a transaction invoked multiple smart-contracts, the policy of each of the {smart-contracts} must be satisfied.
A transaction is marked invalid if it does not have adequate endorsements.
(2) The \texttt{serializability validator} applies optimistic
concurrency control (OCC)~\cite{occ} using the read-write set present in the
transaction response. To facilitate OCC, Fabric adds a version identifier
to each state stored in the blockchain.
The version of a state is stored as $(B, T)$, where $B$ denotes the block number,
and $T$ denotes the transaction number within the block that last updated/created this state.
All the validator checks for is the states, which the transaction has read to decide on its write-set,
have not been modified by any preceding valid transaction.
Further, the validator re-executes range queries present in the \textit{range query info}
to detect any phantom read~\cite{ansi-isolation}. 
\par
Note that the \textit{endorsement policy validator} parallelly validates transactions
in a {block. In contrast,  the \textit{serializability validator} serially validates
each transaction to take into account the writes
of previous valid transactions in the block}.

\begin{figure}
	\begin{minipage}{0.45\textwidth}
		\centering
		\includegraphics[height=60pt]{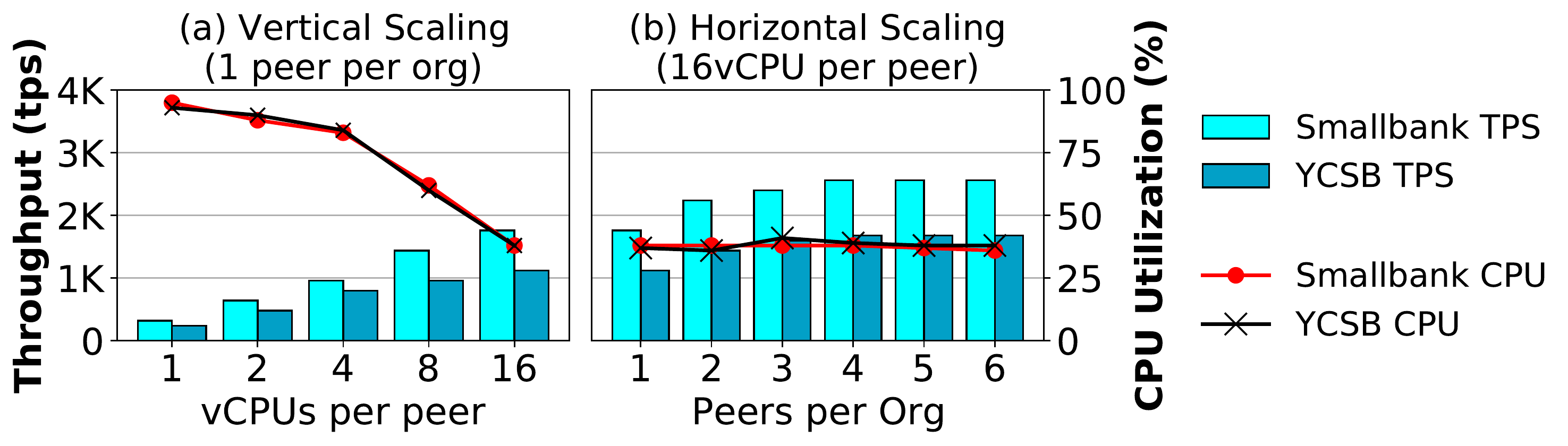}
  		\vspace{-0.4cm}
		\caption{\small{Vertical and horizontal scaling}}
		\label{fig:scaling}
	\end{minipage}
\end{figure}
\begin{figure}
	\begin{minipage}{0.45\textwidth}
		\vspace{-0.5cm}
		\centering
		\includegraphics[height=60pt]{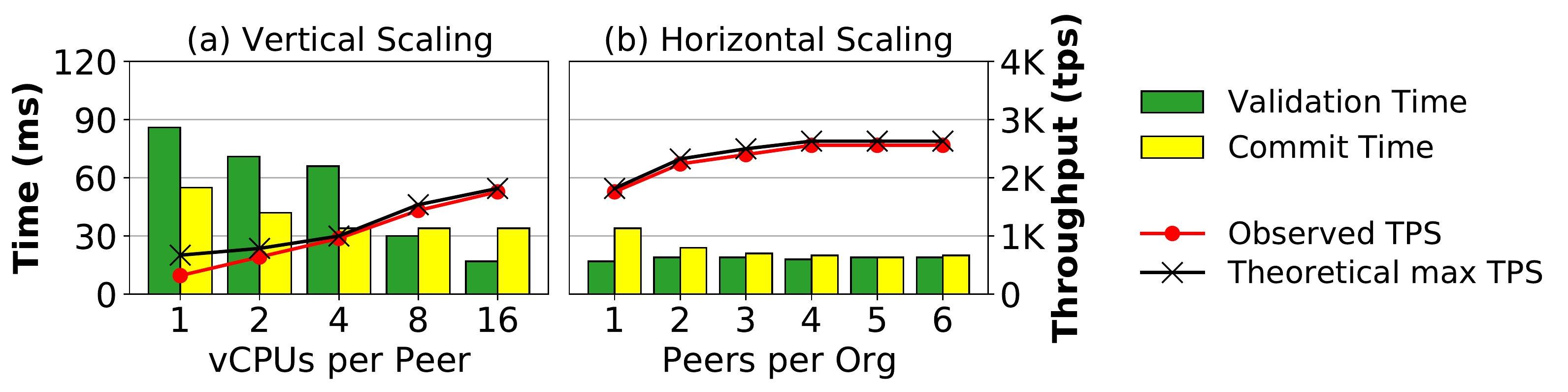}
  		\vspace{-0.8cm}
		\caption{\small{Validation and commit time}} 
		\label{fig:scaling-time}
	\end{minipage}
	\vspace{-0.66cm}
\end{figure}

\par
\textbf{Phase 4---Commit.} After the validation phase,
the \textit{committer} first stores the block in the \textit{block
store}, which is a chain of blocks stored in a file system along with the
transaction validity information. Second, the \textit{committer} applies the write
set of all valid transactions to the \textit{state database}, which maintains
all active states. Third, it stores metadata of updates of all valid transactions to the
\textit{history database}, which maintains the version history (but not the values) of both active and inactive states.
\vspace{-.2cm}
\subsection{Impact of Scaling Techniques on EOV}\label{motivation}
In this section, we quantify the impact of vertical and horizontal scaling on
Hyperledger Fabric's performance.
In vertical scaling,
we add more compute power to the existing nodes. In contrast, we add more nodes in horizontal scaling.
\par
The blockchain network topology used for all experiments consisted of four organizations, each hosting $N$ number of peer
nodes, ordering service based on Raft consensus protocol~\cite{raft} with five nodes,
and $M$ clients to generate load on the network. The value of $M$ and $N$ were changed
depending upon the experiment. Each node was hosted on a
virtual machine with 32 GB RAM, 16 vCPUs of Intel Xeon E5-2683 v3 2.00GHz, SSD storage,
and 1 Gbps of network bandwidth. As the primary performance metric, we measured the transaction throughput,
which is the maximum rate at which a peer commits transactions, beyond which the
peer's block queue would overflow.
Note that our results hold for a network with a large number of organizations too.
This is because the bottleneck identified in this section are not related to the
number of organizations in a network.
\par
\textbf{Workload and Configuration.} Unless specified otherwise, in the default configuration, each organization hosted a peer,
and all organizations are a member of a single channel hosting eight smart-contracts.
All eight smart-contracts either implement the Smallbank~\cite{smallbank} or YCSB\cite{ycsb} benchmark.
The smallbank benchmark consists of 6 operations on a set of 100k accounts. Of the 6, 5 involve both reads
and writes{, and} 1 involves only reads.
To generate load, we have chosen an operation and the accounts uniformly. 
This workload is light on IO as the value's size was 10 bytes. 
The YCSB is a key-value store benchmark with 100K keys,
and each transaction reads and writes two keys. This workload is IO heavy as the value's size was 1 KB.
The keys were chosen using the Zipfian distribution with a skewness factor of 0.5.
The block size was 100 transactions
for all experiments.
\par
\textbf{Base Case Performance.}
Instead of using the vanilla Hyperledger Fabric v1.4 as the base case, we used an optimized
version {that} avoids the redundant block serialization and deserialization at various
phases within a peer using a block serialization cache{, as} proposed in
{FastFabric}~\cite{fastfabric}.
With the block deserialization cache, the throughput improved by 1.67$\times$, i.e.,
from 1040 tps to 1760 tps for the smallbank benchmark using the default configuration.
\par
\textbf{Infrastructure Cost in Public Cloud.}
From cost estimators~\cite{aws-cost, azure-cost, google-cost, ibm-cost} published by public cloud providers,
such as AWS, Azure, Google Cloud, and IBM Cloud,
we identified that the infrastructure cost is a linear function of the number of vCPUs allocated.
For example, a virtual machine (VM) with 8 vCPUs costs almost 8$\times$ of a VM with 1 vCPU.
Similarly, two 8-vCPU VMs cost almost the same as one VM with 16 vCPUs. We use
this trend while explaining the cost of infrastructure in the rest of the paper.
\vspace{-.3cm}
\subsubsection{Vertical Scaling}
Figure~\ref{fig:scaling}(a)
plots the peak throughput and CPU utilization over a different number of allocated vCPU.
With an increase in the allocated vCPUs from 1 to 16, the throughput of the smallbank benchmark
increased disproportionately from 320 tps to 1760 tps while CPU utilization
decreased from 95\% to 40\%. In other words, while the infrastructure cost increased by
16$\times$, the performance improved only by 5.5$\times$.
YCSB showed similar results.
When the number of vCPUs was low, both endorser and validator
contended for the CPU resources, which led to high CPU utilization. With an increase in
the number of vCPUs, the CPU contention reduced, and the \textit{endorsement policy validator}
utilized multiple vCPUs {to validate transactions parallelly}, resulting in higher throughput.
Moreover, during the commit phase (IO heavy) execution, the validation phase
(CPU heavy) does not execute, and vice-versa, which results {in CPU's underutilization}.
This is because the validator cannot validate block ($i+1$) in parallel with the commit
of block ($i$) as the state updates performed during the commit could make the validator
read stale or incomplete state. As mentioned in \S\ref{hf}, the \textit{serializability} validator
validates each transaction serially, further leading to low CPU utilization.


		


\par
Figure~\ref{fig:scaling-time}(a)
plots the time taken by the validation and commit phases to process a block of size 100
transactions under the Smallbank workload.
As expected, the validation time reduced significantly from 86 \textit{ms} to
19 \textit{ms} when the number of vCPUs increased. The commit time also reduced from 55 \textit{ms}
to 36 \textit{ms} when the number of vCPUs increased from 1 to 4 (due to a reduction in CPU contention
between endorser and committer). Beyond 4 vCPUs, no reduction was observed in the commit time.
YCSB shown a similar trend.
\par
Figure~\ref{fig:scaling-time} plots the \textbf{theoretical maximum throughput} and the actual throughput achieved. We define the
theoretical maximum throughput as $ |B| \times \frac{1000\mbox{ }ms}{(V + C)\mbox{ }ms} $ where $|B|$ is the number of
transactions in block $B$, $V$ is the validation time of block B, and $C$ is the commit time of block B. In other words,
it is equal to the maximum number of transactions that can be validated and committed in a second.
When there was only 1 vCPU, the actual throughput was much lower than the theoretical maximum.
This is because the endorser occupied some CPU cycles that made the validator/committer wait.
\par
\textbf{{\hypertarget{takeaway1}{\underline{Takeaway 1}.}}} \textit{Vertical scaling of
		peers can help to a degree. However, it is necessary to have pipelined execution of validation
		and commit phases to utilize the full potential of allocated vCPUs and get a better
        return on investment.}
\vspace{-.1cm}
\subsubsection{Horizontal Scaling}
Figure~\ref{fig:scaling}(b)
plots the throughput and CPU utilization over many
peers per organization. Compared to 1 peer per organization,
4 peers increased the throughput by only 1.5$\times$ (for both workloads) while the infrastructure cost
increased by 4$\times$. This is because all 4 peers within
each organization validated and committed each block, i.e., redundant work within an organization.
With an increase in the number of peers, only the endorsement for transactions {was load-balanced}
between peers by clients, which also reduced the CPU utilization. As the
validation and commit phases are costlier than the simulation phase due to multiple signature
verifications and synchronous disk IO, the throughput increment is not significant.
Beyond 4 peers per organization, the throughput did not increase.
As load balancing of endorsement requests increased the
throughput, we wanted to {find the peer's peak throughput
with zero endorsement requests}.
Hence, we did not submit any endorsement requests for a peer while using other
peers to load balance the endorsement requests. We observed that the non-endorsing peer
achieved a peak throughput of 3300 tps.
Figures~\ref{fig:scaling-time}(b) plots the validation and commit time 
for smallbank. 
With an increase in the number of peers, the validation time almost stayed the same.
However, the commit time reduced dramatically from 34 \textit{ms} to 20 \textit{ms}
due to a reduction in both CPU \& IO contention between the endorser and committer. 
\par
\textbf{\hypertarget{takeaway2}{\underline{Takeaway 2.}}} \textit{Horizontal scaling of a Fabric network by adding more
		peers
can help {reduce} the endorser's load, i.e., {the simulation phase. Still, it}
does not help the
validation and commit phases much due to redundant work. Further, it does not justify the additional
infrastructure cost paid for the performance gain. Redundant work is needed
across organizations as one organization does not trust another. However, the same does not apply
to peers within an organization. It is necessary to avoid redundant
work within an organization to improve performance and reduce cost.}
\par
\begin{figure}[t]
    \begin{center}
			\includegraphics[scale=0.4]{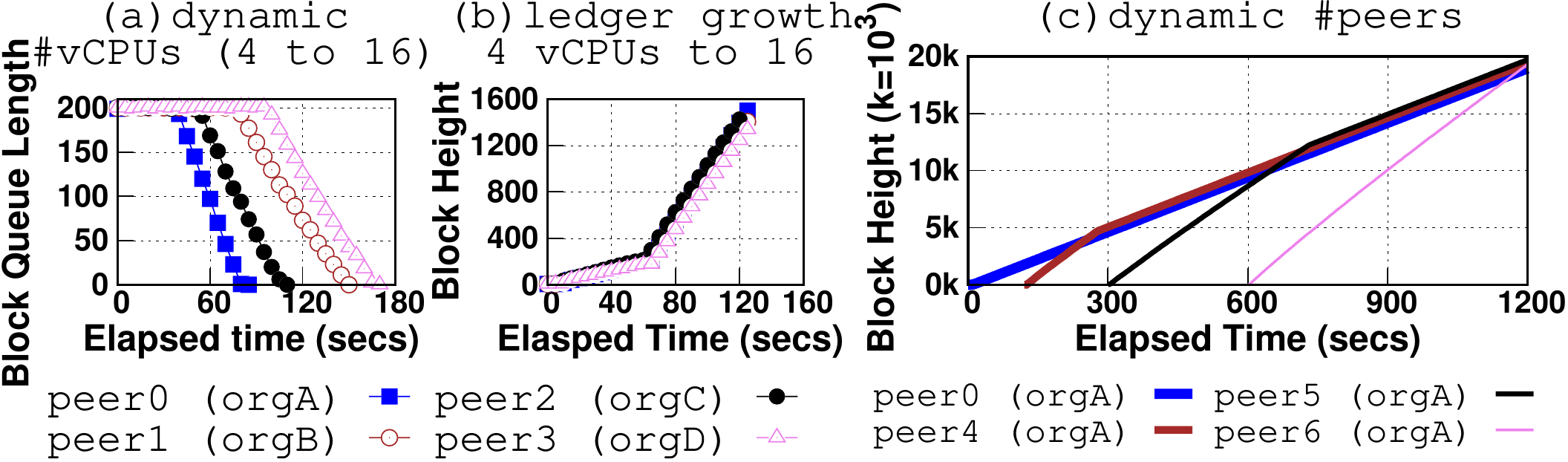}
  		\vspace{-.6cm}
		\caption{\small{Impact of dynamic scaling by vCPUs \& peers.}}
		\label{fig:dynamic}
 \end{center}
  \vspace{-.6cm}
\end{figure}
\vspace{-.2cm}
\subsection{Mitigation of an Overloaded Situation}
\par
\subsubsection{Vertical Scaling}
To study the impact of dynamic
vCPU scaling on the performance, we overloaded peers in a network by generating more
load than it can handle at 4 vCPUs. We then increased the number of vCPUs to
16, one peer at a time, after the block queue's length reached 200.
Figure~\ref{fig:dynamic}(a) plots the time taken to reduce the block queue length
from 200 to 0 for each peer across organizations. Though peers scaled
immediately, it took around 50 \textit{secs} to 70 \textit{secs} to reduce the
queue length to 0.
Figure~\ref{fig:dynamic}(b) plots the ledger block height over time, while the
number of vCPUs increased from 4 to 16. The block height is nothing but the
last committed block number. As expected, the ledger commit rate increased.
Though the queue size became 0 within 70 \textit{secs} and the ledger commit rate
increased, there was a significant impact on the number of failed transactions
\begin{wrapfigure}{r}{0.14\textwidth}
  \vspace{-.4cm}
  \begin{center}
		  \includegraphics[width=0.15\textwidth]{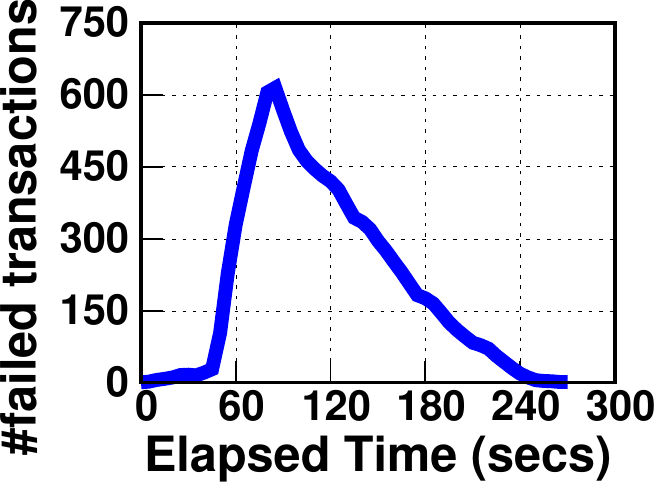}
  \end{center}
  \vspace{-.3cm}
  \caption{\small{Failed transactions.}}
  \label{fig:failed}
  \vspace{-.4cm}
\end{wrapfigure}
due to serializability violations (as shown in Figure~\ref{fig:failed}). As more blocks were
waiting in the queue to update the ledger state, new transactions were endorsed using
very stale states. Thus, many transactions in each block were invalidated later
by the \textit{serializability validator} due to a mismatch in the state's version
present in the read set against the committed ledger state. After scaling peers,
it took 180 seconds to reduce the number of failed transactions to < 2.
\par
\textbf{\hypertarget{takeaway3}{\underline{Takeaway 3.}}} \textit{Dynamic scaling by vCPU is efficient as it
can quickly react to the increased load. However, this approach is limited by
the number of available vCPUs}. 
\vspace{-.1cm}
\subsubsection{Horizontal Scaling}
\label{hs-mitigation}
To study the time taken to add
a new peer in an existing organization, we ran a
peer per organization and then added a new peer at the $2^{nd}$ minute, $5^{th}$
minute, and $10^{th}$ minute. Figure~\ref{fig:dynamic}(c) plots the time
taken by new peers to sync up with the existing peers to
process new endorsement requests and new blocks.
We observed that the time taken was proportional to existing peers' block height
as the new peer had to fetch all old blocks, validate, and commit them one by one.
As there
were no endorsement requests on the new peer during the sync up,
it could catch up with the existing peer by committing transactions
at a rate of 3300 tps (while with endorsement, it was only 1760 tps).
\par
\textbf{\hypertarget{takeaway4}{\underline{Takeaway 4.}}} \textit{
With the dynamic scaling by
peers, the new peer took a significant amount of time to sync up with existing
peers to become available for handling the load. This is
because the new peer validated and committed all blocks, i.e.,
redundant work.}

%% file: design.tex
\vspace{-.3cm}
\section{Design of Pipelined Execution}
\begin{figure}
  \centerline{\includegraphics[scale=0.31]{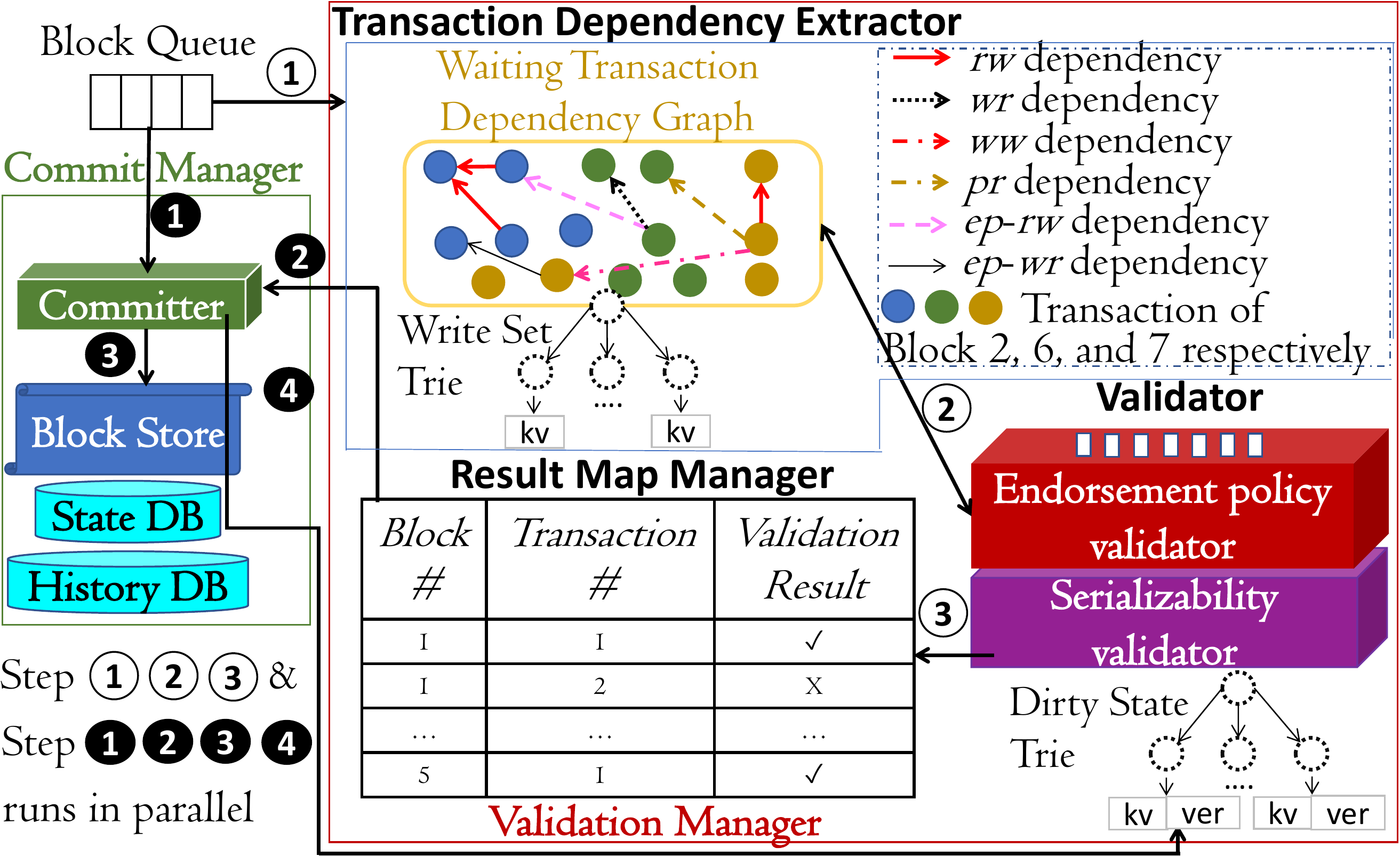}}
  \vspace{-.3cm}
  \caption{\small{Proposed architecture to enable pipelined execution of validation and commit phases.}}
  \label{fig:eagervalidator}
  \vspace{-.5cm}
\end{figure}
\label{parallel-validation}
As mentioned in \hyperlink{takeaway1}{takeaway 1} in \S\ref{motivation}, we need to increase the CPU
utilization. We can do that by making the validator (i.e., CPU heavy) validate block ($i+1$) without waiting for
the committer (i.e., IO heavy) to commit block $i$. Besides, we can validate transactions in parallel (even across blocks).
For the former, the challenge is to ensure that the \textit{validator is not reading any stale state}. For
the latter, the challenge is to ensure that the \textit{serializable schedule is equivalent to the
transactions ordering in blocks}. Towards achieving this, we propose a new architecture shown in
Figure~\ref{fig:eagervalidator}.
The two fundamental ideas of the proposed architecture are as follows:
\begin{enumerate}[topsep=2pt,itemsep=0pt,leftmargin=15pt]
    \item Exploit the \textit{read-write set} and \textit{range query info} present in each transaction to construct a dependency graph. This graph helps to choose transactions for parallel validation without violating serializability.
    \item Maintain a dirty state buffer to keep all uncommitted writes of valid transactions.
    This buffer helps avoid reading a stale state while pipelining validation and commit phases.
    Thus, the validator can validate a future block without waiting for past blocks to get committed.
\end{enumerate}
\vspace{-.4cm}
\subsection{Validation Phase} 
The proposed validation phase consists of three components: (1) transaction dependency extractor, (2) validator, and (3) a result map manager, as shown in Figure~\ref{fig:eagervalidator}.
In step {\ding{172}}, the extractor reads a block from the queue and updates the dependency graph.
In step \ding{173}, each free validator worker performs the following:
(i) retrieve a transaction from the graph that has no out-edges \& validates it;
(ii) if validation passes, apply the write-set to dirty state;
(iii) remove the transaction from the graph.
In step \ding{174}, the worker adds the validation result to the \textit{result map} before moving to step \ding{173}.
\par
\vspace{-.2cm}
\subsubsection{Transaction dependency extractor}
The extractor adds each transaction in a block to the \textit{waiting-transactions dependency graph}.
This graph contains a node per transaction. 
We interchangeably use node and transaction.
Let us assume $T_{i}$ and $T_{j}$ are two transactions such that $T_{i}$ appeared before $T_{j}$ ($T_{j}$ can be in the same block or any later block).
An edge from $T_{j}$ to $T_{i}$ (i.e., $T_{i} \leftarrow T_{j}$) denotes that $T_{i}$ must be validated before validating $T_{j}$.
Note that a dependency edge is always from $T_{j}$ to $T_{i}$ because $T_{i}$ appeared earlier than $T_{j}$.
This ensures that the \textit{serializable schedule is the same as the transaction order present in the block}.
The following are the seven dependencies by which $T_{j}$ can depend upon $T_{i}$:
 \begin{enumerate}[topsep=0pt,itemsep=-4pt,leftmargin=12pt]
     \item \textbf{\textit{read-write}} ($T_{i}\xleftarrow{rw\mbox{(}k\mbox{)}}T_{j}$):
            $T_{i}$ writes a new value to the state $k$, and $T_{j}$ reads a previous version of that state;
     \item \textbf{\textit{write-read}} ($T_{i}\xleftarrow{wr\mbox{(}k\mbox{)}}T_{j}$):
            $T_{j}$ writes a new value to the state $k$, and $T_{i}$ reads a previous version of that state;
     \item \textbf{\textit{write-write}} ($T_{i}\xleftarrow{ww\mbox{(}k\mbox{)}}T_{j}$):
            Both $T_{i}$ and $T_{j}$ write to the same state $k$;
     \item \textbf{\textit{phantom-read}} ($T_{i}\xleftarrow{pr\mbox{(}k\mbox{)}}T_{j}$):
            $T_{j}$ performs a range query, and $T_{i}$ creates a new or deletes an existing state $k$ that would match that range query;
     \item \textbf{\textit{endorsement-policy read-write}} ($T_{i}\xleftarrow{ep-rw\mbox{(}s\mbox{)}}T_{j}$): 
            $T_{i}$ updates the endorsement policy of a smart-contract $s$, and $T_{j}$ invokes $s$ using a previous version of the policy;
     \item \textbf{\textit{endorsement-policy write-read}} ($T_{i}\xleftarrow{ep-wr\mbox{(}s\mbox{)}}T_{j}$): 
            $T_{j}$ updates the endorsement policy of a smart-contract $s$, and $T_{i}$ invokes $s$ using a previous version of the policy;
     \item \textbf{\textit{endorsement-policy write-write}} ($T_{i}\xleftarrow{ep-ww\mbox{(}s\mbox{)}}T_{j}$): 
            Both $T_{i}$ and $T_{j}$ update the endorsement policy of the same smart-contract $s$;
 \end{enumerate}
No cycle is possible in the graph as an edge is always from a new transaction to an old one.
\par
\textbf{{Fate dependencies}:} While all seven dependencies are necessary to choose a set of transactions for
parallel validation, the following three dependencies also decide the validity of dependent transactions:
\begin{enumerate}[topsep=0pt,itemsep=0pt,leftmargin=30pt]
    \item read-write (\textit{rw}) dependency
    \item phantom-read (\textit{pr}) dependency
    \item endorsement policy read-write (\textit{ep-rw}) dependency
\end{enumerate}



\begin{figure*}[th]
    \begin{minipage}{0.53\textwidth}
        \vspace{-.5cm}
        \captionof{table}{Read-write sets of transactions.
        \vspace{-.45cm}
        \label{table:eagerval-illustration-rwset}} 
        \raggedright
        \input{tables/dependency-graph}
    \end{minipage}
    \hspace{0.05cm}
    \begin{minipage}{0.40\textwidth}
        \vspace{-.5cm}
        \centering
        \caption{Dependency graph of transactions} 
        \vspace{-.1cm}
        \includegraphics[scale=0.6]{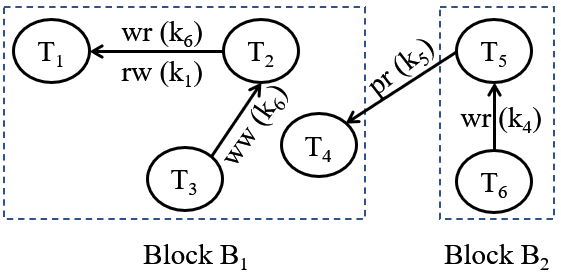} 
        \label{fig:dependency-graph}
    \end{minipage}
\end{figure*}
\begin{table*}
    \vspace{-.2cm}
    \centering
    \vspace{-0.35cm}
    \caption{Validation process of transactions}
        \vspace{-.5cm}
\input{tables/illus-validator.tex}
    \label{table:illus-val}
    \vspace{-0.3cm}
\end{table*}

Hence, these dependencies are called \textit{fate dependencies}.
When there is a \textit{fate dependency} from $T_{j}$ to $T_{i}$, i.e., $T_i \xleftarrow{rw\mbox{/}pr\mbox{/}ep-rw}
T_j$, and if $T_{i}$ turns out to be valid, $T_{j}$ must be invalid. For instance, if $T_{i}$ modified a state
$k$ so that its new version is $10$, but $T_{j}$ had read $k$ at version $9$ (\textit{rw} dependency), $T_{i}$ being
valid implies that $T_{j}$ is invalid.
For any other dependencies, $T_{j}$ can be valid or invalid irrespective of the validation result of $T_{i}$.
For instance, if $T_{i}$ read $k$ at version $9$ and $T_{j}$ modified it to version $10$ (\textit{wr} dependency),
the dependency only ensures that $T_{j}$ does not modify the state before $T_{i}$ is validated. However, both can
be valid or invalid independently.

\par

\textbf{Detecting dependencies:}
To track dependencies, we maintain a map from state $k$ to ($T_{R}^{k}$, $T_{W}^{k}$), where $T_{R}^{k}$ and $T_{W}^{k}$ are
transactions in the graph that respectively read and write to $k$.
When adding a new transaction, we lookup $(T_R^k, T_W^k)$ for every state $k$ it reads or writes and add appropriate dependencies.
When adding a transaction $T_j$ that performs a range query, we need to find writers of every state in that range to detect $pr$ dependencies.
Simply iterating through the range and looking up every state in the map would not be feasible because the ranges could be large or unbounded.
Therefore, we store the keys internally as an ordered tree (e.g., a Trie) to perform both point and range queries quickly.
Thus, by running $T_j$'s range query on this tree, we can efficiently find all states in the range that are being modified by some waiting transactions.
We can then add \textit{pr} edges from $T_j$ to writers of those states.


\par

\textbf{Illustration of dependency extraction}.
Let us assume at the start of a peer, blocks $B_{1}$ and $B_{2}$, listed in Table~\ref{table:eagerval-illustration-rwset}, are waiting in the queue.
Note that an entry in the write-set also includes the new value, but we do not present it for brevity. $v_1 \rightarrow v_2$ denotes that the version is being incremented.
The dependency extractor would fetch block $B_{1}$ to start constructing the dependency graph.
Transaction $T_{1}$ gets added as the first node in the graph.
$T_2$ then gets added with two dependencies on $T_1$: (a) \textit{rw} dependency because $T_2$ reads state $k_{1}$
while $T_1$ updates it and (b) \textit{wr} dependency because $T_2$ updates $k_6$ but $T_1$ reads it.
Similarly, a \textit{ww} dependency gets added from $T_{3}$ to $T_{2}$, as shown in Figure~\ref{fig:dependency-graph}.
The fourth transaction in block $B_{1}$ has no dependency on other transactions in the same block.
Once the dependency extractor adds all transactions in block $B_{1}$ to the graph, it would fetch block $B_{2}$ and add each transaction to the graph.
Transaction $T_{5}$ has a phantom read dependency on $T_{4}$ because the former queries range $[k_2, k_5]$ while the latter
updates $k_{5}$ that belongs to the range.
Transaction $T_{6}$ has a \textit{wr} dependency on $T_{5}$ because it updates $k_{4}$ while $T_{5}$ reads the previous
version of the same key using a range query.
\par
\textbf{Operations exposed to other components:}
The extractor exposes the following two operations to enable other components to access the dependency graph:
(1) \texttt{\small{GetNextTransaction() $\rightarrow T_{i}$}}: Returns a reference to the oldest transaction $T_i$ in the
        graph that does not depend on any other transactions (but other transactions can depend on it). 
(2) \texttt{\small{RemoveFromGraph($T_{i}$, isValid)}}: Removes $T_i$ from the dependency graph.
        If \texttt{\small{isValid}} is \textit{true}, i.e., $T_{i}$ is valid, it marks all transactions
        depending on $T_{i}$ via a \textit{fate dependency} as invalid and removes them from the graph.
\par
\vspace{-.2cm}
\subsubsection{Validators}
The logic of \texttt{\small{endorsement policy validator}} and
\texttt{\small{serializability validator}} remains the same as the one discussed in \S\ref{hf}
except the following modifications:
\begin{enumerate}[topsep=1pt,itemsep=-0pt,leftmargin=15pt]
        \item Validators can validate later blocks without waiting for the committer to commit earlier blocks.
        \item Validators read from dirty state buffer before resorting to state DB to avoid reading stale states.
        \item Multiple workers to validate transactions in parallel.
\end{enumerate}
\par
Each free worker calls \texttt{\small{GetNextTransaction()}} to get the next transaction to be processed.
First, the \texttt{\small{endorsement policy validator}} checks whether the policy is satisfied.
On success, the same worker executes the \textit{serializability} check using OCC~\cite{occ}. If the transaction passes both validations,
the worker first applies the write-set to the dirty state and then calls \texttt{\small{RemoveFromGraph()}} to
update the dependency graph. Finally, the worker adds the validation result to the \textit{result map}. To ensure
that the validator does not read a stale state when it validates block j before committing block i (where j > i),
a read request (such as a read of the endorsement policy or the version of a state) would first go to the dirty state.
Only on a miss, the read would reach the state DB. We store the dirty state as a trie, with the state's key
stored along a path and its value \& version stored in a leaf. This trie structure enables the
\textit{serializability validator} to validate range queries present in \textit{range query info} for a phantom read.
Every range query is executed on both the trie and state DB. 
\par
\textbf{Illustration of Validators}.
Let us assume we have three workers, $w_1$, $w_2$, and $w_3$, to validate transactions, as shown in Table~\ref{table:illus-val} (only the first 3 rows are relevant for this illustration).
Workers $w_1$ and $w_2$ would get transaction $T_1$ and $T_4$, respectively, by calling \texttt{\small{GetNextTransaction()}}.
As every other transaction in the dependency graph has an out-edge, the $w_3$ would stay idle.
Suppose the state database initially has three states, as shown in the first row of Table~\ref{table:illus-val}.
For simplicity, let us further assume workers $w_1$ and $w_2$ would always find endorsement policy to be satisfied.
In other words, the \textit{endorsement policy validator} would always return success.
\par
Once the endorsement policy check succeeds, the worker would perform the \textit{serializability} check in which the state's version
present in the read-set of the transaction should match the respective state's version present in the dirty state or state DB.
As per OCC, if there is a mismatch, the transaction would be marked invalid.
As workers $w_1$ and $w_2$ find the respective transaction to pass the serializability check, both transactions would be marked as valid.
Each worker then applies its transaction's write-set to the dirty state (refer to the second row in Table~\ref{table:illus-val}).
Finally, the workers would call \texttt{\small{RemoveFromGraph()}} to update the dependency graph.
\par
Given that transaction $T_2$ has a fate dependency on transaction $T_1$, $T_2$ would be marked invalid because $T_{1}$ is valid.
Similarly, since transaction $T_5$ has a fate dependency on transaction $T_4$, $T_5$ would be marked invalid because $T_4$ is valid.
Thus, the dependency graph would have only two transactions ($T_3$ and $T_6$) left, and they would have no edges between them.
Next, the worker $w_1$ would pick $T_3$, and $w_2$ would pick $T_6$.
As per OCC, both the transactions would be valid, and the dirty state gets updated with their write sets.
On calling \texttt{\small{RemoveFromGraph()}}, the dependency graph would be left empty.
Note that only after the committer commits a block would the entries in the dirty state be removed.
\vspace{-.2cm}
\subsubsection{Result map manager}
This manages a map from transaction identifier to its validation result. It exposes the following two operations:
    (1) \texttt{\small{AddValidationResult}} ($T_i$, \textit{isValid}) adds the validation result of a transaction to the map---called
    by both \textit{extractor} and \textit{validator}.
    (2) \texttt{\small{PopValidationResult}}($T_i$) returns the validation result associated with a given transaction after deleting it---called
    by the \textit{committer}.
\vspace{-.2cm}
\subsection{Commit Phase} 
The logic of the commit phase does not change significantly as compared to the vanilla Fabric discussed in \S\ref{hf}.
As shown in
Figure~\ref{fig:eagervalidator}, in step \ding{182}, whenever the \textit{committer} becomes free, it reads a block
from the queue and retrieves the list of transactions. In step \ding{183}, the committer fetches
the validation results by calling \texttt{\small{PopValidationResult()}} for each transaction. If the validation result is
not available for a transaction, the call would be blocked until the result is available. Once the committer
collects {the} validation result of all transactions, in step \ding{184}, it stores the block in the block store and applies
the valid write-sets to state \& history DB as in vanilla Fabric. In step \ding{185}, it calls the validation
manager to remove the dirty state associated with the just committed block as the validator can read
those states from the state DB itself.
\par
\textbf{Illustration of Committer}.
Let us continue with the scenario listed in Table~\ref{table:illus-val}. In parallel with the validation manager, the
committer would fetch block $B_1$ and wait for the validation result of $T_1$, $T_2$, $T_3$, and $T_4$. Once these validation results
are available, the committer would commit $B_1$, as shown in Table~\ref{table:illus-val}.
Then, the dirty state and results associated with the block would be deleted. Next, the committer would fetch
block $B_2$ and immediately find validation results for all transactions. Hence, the committer would commit it and update the dirty state and result list.
\vspace{-.2cm}
\subsection{Proof of correctness}
{
The ordering service predetermines the commit order of transactions.
Therefore, we have to ensure that our approach validates transactions such that
it is equivalent to the predecided serial schedule.
Let us say the order of transactions is $T_1, T_2, \cdots, T_n$ in a block. 
To validate a transaction, say $T_{5}$, the validator checks that all states read by $T_5$
have not been modified by any previous transactions (present in the same or previous blocks).
If the last transaction that updated any of those states was $T_1$,
we do not need to wait for the validation result of any transaction between them to validate $T_5$.
We only need to ensure that if $T_1$ is valid, the validator sees $T_1$'s updates while validating $T_5$.
This requirement is easily met by \textit{rw}-dependencies.
Since $T_5$ has an \textit{rw}-dependency on $T_1$, the validator workers will never get $T_5$ before $T_1$ is removed from the graph.
Once a validator worker receives $T_1$ via \texttt{GetNextTransaction()}, it first validates $T_1$.
If valid, it applies $T_1$'s updates to the dirty state.
Only then does it call \texttt{RemoveFromGraph($T_1$)}.
Note that to process any transaction, all transactions it depends upon must be removed from the graph.
Thus, when a worker receives $T_5$, $T_1$'s updates would be either in the
dirty state (if $T_1$ is valid but not yet committed) or in State DB (if $T_1$ is valid and committed).
Thus, the worker can correctly validate $T_5$. If $T_1$ is valid, $T_5$ would be invalid. Instead of
wasting the validator resources, we term \textit{rw} dependency as \textit{fate dependency}
and immediately invalidate the dependent transaction if the dependee is valid.

However, \textit{rw}-dependencies are not sufficient.
Consider transactions $T_i$ and $T_j$ ($i<j$) in block $B_1$ that both write to the same state.
Without \textit{ww}-dependency, $T_i$ updates the dirty state after $T_j$, overwriting $T_j$'s update in the dirty state.
Now, let's say that before $B_1$ is committed, this peer started validating the next block having a transaction $T_k$ that has read the state written by $T_j$.
This could happen if $T_k$ was endorsed on a peer where $B_1$ was already committed (e.g., on a much faster peer, or if this peer has just joined the network).
Now, the validator would wrongly mark $T_k$ as invalid because it would see $T_i$'s write in the dirty state while expecting $T_j$'s write.
To avoid such anomalies, we track \textit{ww} and \textit{wr} and ensure that updates to a state are applied in the correct order.
The same argument can be extended to ep-rw, ep-wr, and ep-ww dependencies. Phantom read is just a particular case of \textit{rw}-dependency.}
\vspace{-.2cm}
\section{Design of Sparse Peer}
\label{sparse-peer}
From \hyperlink{takeaway2}{takeaway 2} in \S\ref{motivation}, we know that multiple peers {in an organization do redundant work,
limiting horizontal scaling efficiency}. To avoid redundancy, we propose a
new peer type called a \textit{sparse peer}.
{A \textit{sparse peer} may not validate and commit all transactions within a block
compared to a full peer in vanilla Fabric. The sharding concept in a distributed database inspires the concept of a \textit{sparse peer}.}
\vspace{-.2cm}
\subsection{Sparseness in Validation and Commit} 
\label{sparseness}
The key idea behind a \textit{sparse peer} is that it can selectively validate and commit
transactions. If all \textit{sparse peers} within an organization select a non-overlapping set of transactions,
we can avoid redundant work. Towards achieving
this, first, we define a deterministic selection logic such that each \textit{sparse peer} selects a
different set of transactions. Second, we change the validator and committer
to apply the selection logic on the received block. Third, as an optional feature, we make the peer pass
the selection logic to the orderer such that the orderer itself can apply the filter and send only required
transactions in a \textit{sparse block}. Thus, both network bandwidth utilization and disk IO would reduce.
\begin{figure}
\centerline{\includegraphics[scale=0.3]{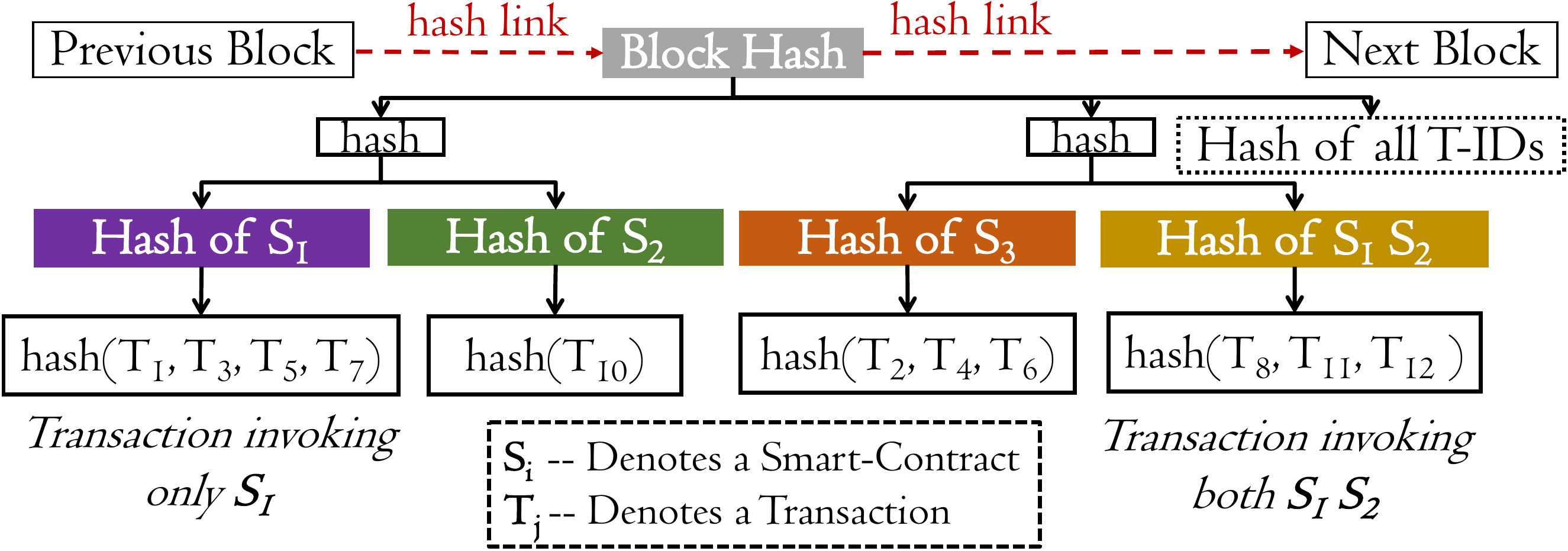}}
\vspace{-.4cm}
\caption{\small{Merkle tree based block hash computation.}}
\label{fig:sparseIndex}
\vspace{-.5cm}
\end{figure}
\par
\textbf{{(1) Transaction selection filter}}: Each sparse peer owns a \textit{filter} and applies it
on a received block to identify which transactions to consider. The \textit{filter} is
simply a list of smart-contracts.
The admin assigns/updates each peer's \textit{filter} by issuing a request via gRPC to the peer process.
The \textit{sparse peer} only validates and commits transactions in a block that invoked a smart-contract
specified in the \textit{filter}. When a transaction had invoked multiple smart-contracts, even if
the filter contains only one of those smart-contracts, the transaction would be considered by the
\textit{sparse peer}.
\begin{figure*}
    \hspace{-.0cm}
    \begin{minipage}{0.27\textwidth}
        \vspace{-.2cm}
        \captionof{table}{Transaction RW-sets}
        \vspace{-.5cm}
        \centering
        \label{fig:eagerval-illustration-rwset}
        \input{tables/transactions-sparse-peer.tex}
    \end{minipage}
    \hspace{0.25cm}
    \begin{minipage}{0.27\textwidth}
        \raggedleft
        \vspace{-.8cm}
        \captionof{table}{StateDB at each peer\label{table:sparse-peer-statedb}} 
        \vspace{-.5cm}
        \centering
        \input{tables/statedb-sparse-peer.tex}

    \end{minipage}
    \begin{minipage}{0.42\textwidth}
        \vspace{-.5cm}
        \centering
        \caption{Transaction dependency graph} 
        \hspace{-0.2cm}\includegraphics[scale=0.40]{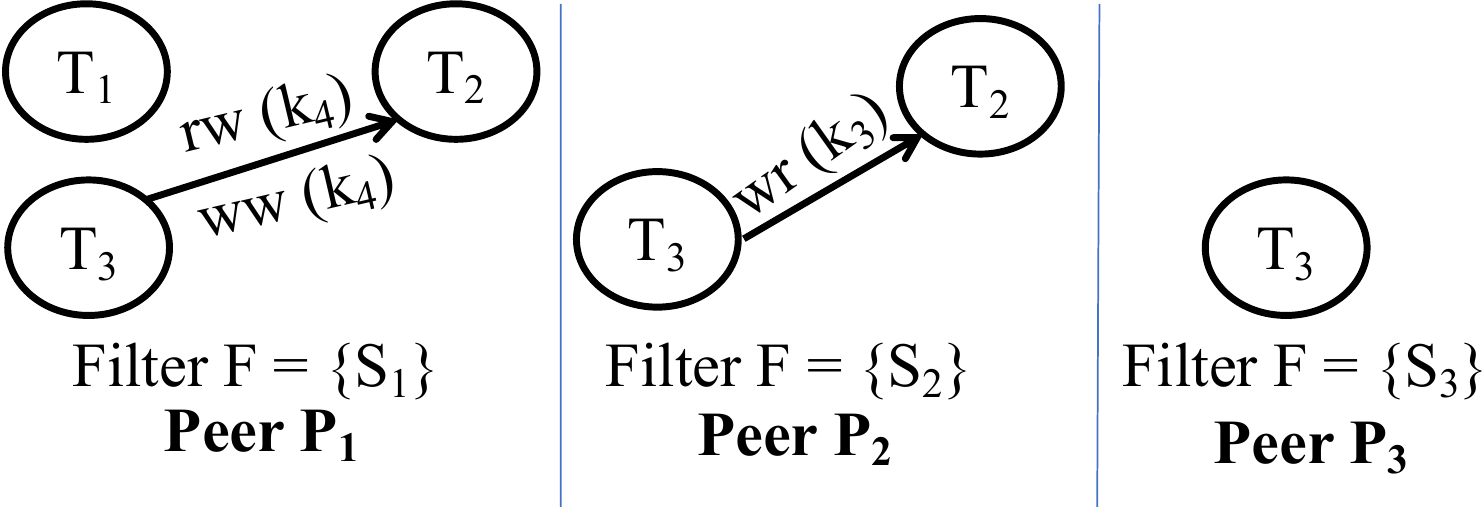} 
        \label{fig:sparse-peer-dependency-graph}
    \end{minipage}
\end{figure*}
\begin{table*}
    \centering
    \vspace{-0.4cm}
    \caption{Distributed validation of transactions}
    \vspace{-0.4cm}

\input{tables/illus-sparse-peer.tex}
    \label{table:illus-val-sparse-peer}
    \vspace{-0.3cm}
\end{table*}
\par
\textbf{(2) Validation and commit based on a filter.}
The dependency extractor only considers transactions that invoked a smart-contract present in the \textit{filter}.
When the committer reads the block, it marks transactions that did not
invoke any smart-contract present in the \textit{filter} as not validated but stores the whole block
in the block store. The rest of the validator and committer logic remains the same.
However, the receival and storage of full block would not reduce network bandwidth utilization and disk IO.
Since disk IO is in the critical path of peers, this limits the maximum possible throughput.
As an optional feature, next, we allow sparse peers to pass their filter directly to the orderer.
\par
\textbf{(3) Block dissemination based on filters.} If orderers themselves apply the filter and send
only appropriate transactions via a \textit{sparse block} to each \textit{sparse peer}, we can save both network bandwidth utilization and
disk IO. Hence, each \textit{sparse peer} sends its filter to an orderer to which it has connected.
For each block, the orderer applies the filter and sends only the
required transactions to the peer. However, this creates a problem with the hash chain and its
verification. In vanilla Fabric, the orderer computes a \textit{block hash} at block creation
and stores it in the block. The \textit{block hash} is computed using all transactions' bytes within that
block and the hash present in the previous block. When a peer receives a block, it can check its
integrity by verifying the hash chain. Further, this hash chain is the source of truth of a blockchain
network. If we make the orderer send only a sub-set of transactions in a block, the peer cannot
verify the hash chain integrity.
\par
\textbf{\textit{Sparse block.}} To fix this problem, we propose a \textit{sparse block} that includes (1) a Merkle tree to represent
the \textit{block hash} (as shown in Figure~\ref{fig:sparseIndex}); (2) only a sub-set of transactions
after applying the filter; (3) applied filter; (4) {transaction identifier (T-ID) of each transaction}.
In our Raft-based consensus, the leader node constructs the {Merkle} tree while
{followers apply the} filter before sending the block to its connected peers.
When a \textit{sparse peer} receives a \textit{sparse block}, it can verify the hash chain integrity
by verifying the Merkle tree root hash using a sub-set of transactions. 
A list of transaction identifiers {is} sent with a \textit{sparse block} to enable the
validator to check for duplicates and mark them invalid.
In vanilla Fabric, the orderer does not peek into the transaction.
We break that as the orderer needs to find transactions associated with each smart-contract and 
find transactions {that} invoke multiple smart-contracts. Since the orderers already have access to the entire
transaction, even in a vanilla Fabric, a motivated party could always peak into the transactions.
Hence, our approach does not {weaken} the trust model. 
\vspace{-.2cm}
\subsection{Distributed Simulation} 
\label{distributed-simulation}
In vanilla Fabric, a smart-contract can invoke another contract hosted on the same channel on the same peer. With sparse peers,
smart-contracts would be placed on different peers. Hence, we enable distributed simulation in which a smart-contract
hosted on one peer can invoke a contract hosted on another by making a gRPC request.
The \textit{filter DB} holds each \textit{sparse peer}'s filter and is used to
decide which peer to contact for a given contract. As the distributed simulation happens
over a network and the \textit{endorser} holds a read lock on the whole state DB~\cite{perf-mascots}, the commit operation
would get delayed if there are many distributed simulations. Hence, we adopt the technique proposed
by Meir et al.~\cite{perf-concurrency} to {remove} the state DB's read-write lock.

\vspace{-.25cm}
\subsection{Distributed Validation and Commit} 
\label{distributed-commit}
Due to the distributed simulation, now we need distributed validation and commit.
Consider a transaction $T$ invoking smart-contracts $S_1$, $S_2$, \dots $S_n$. Consider
sparse peers $P_1$, $P_2$, \dots, $P_n$ where each peer $P_i$ has filter $F_i$ with a single smart-contract
$S_{i}$. The transaction invoking all $n$ smart-contracts is considered valid when the transaction
satisfies both the policy and serialization checks of each contract invocation. Hence, to
commit this transaction, we would require an agreement from all $n$ \textit{sparse peers} during the
validation phase.
\par
\textbf{Strawman protocol.} Each peer $P_{i}$ 
validates parts of the transaction that involved smart contracts $S_i \epsilon F_i$, and then it broadcasts the results
to every other peer. Once a peer has received \textit{valid} as a result for all smart-contracts $S_1 \dots S_k$,
it can consider the transaction valid and commit it. If an \textit{invalid} result is received, the peer does not
need to wait for any more results and can proceed by invalidating the transaction. As peers within an
organization are trusted, there are no security issues. This approach has a significant
drawback. Since the peers could be at different block heights due to different block commit rates (as a result
of heterogeneous hardware or workloads), a transaction requiring distributed commit could block
other transactions in the block for a considerable amount of time. As a result, the \textit{committer}
could get blocked at \texttt{\small{PopValidationResult()}} call.
\par
\textbf{An Improved protocol using deferred transactions.}
To avoid the \textit{committer} getting blocked, we introduce deferred transactions. As a result, the committer
can commit all local transactions (i.e., the partial block commit) without waiting for
distributed transactions' validation results. Further, the committer would mark and store distributed transactions as deferred
during the commit (as it is needed for recovery after a peer failure). Whenever the result is available, the
deferred transaction is committed and removed from the graph. Note that any local transaction with a
dependency on a deferred transaction must be deferred even if it is in a different block.
\vspace{-.3cm}
\subsubsection{Illustration of Distributed Validation}
Let us assume we have three sparse peers, $P_1$, $P_2$, and $P_3$, in an organization. Each peer has
only one smart-contract in the filter. Let us assume each peer has received a full block with three
transactions, as listed in Table~\ref{table:rwset-sparse-peer}. Table~\ref{table:statedb-sparse-peer}
presents the current committed state at each peer's state database.
On each peer, the dependency extractor would apply the filter and construct the dependency graph
shown in Figure~\ref{fig:sparse-peer-dependency-graph}.
As peer $P_1$ has smart-contract $S_1$ in its filter, all three transactions are added while $P_2$
and $P_3$ have only $T_2$ \& $T_3$ and $T_3$ in the dependency graph, respectively. Note that even
though $T_3$ has read from and writes to all three smart-contracts, peer $P_2$ would consider
only $S_2$-states in the read-write set while constructing the dependency graph. The same is true
for the other two peers. This does not affect the correctness, as the example below shows.
\begin{figure*}[t]
	\begin{minipage}{0.57\textwidth}
		\raggedright
		\includegraphics[height=85pt]{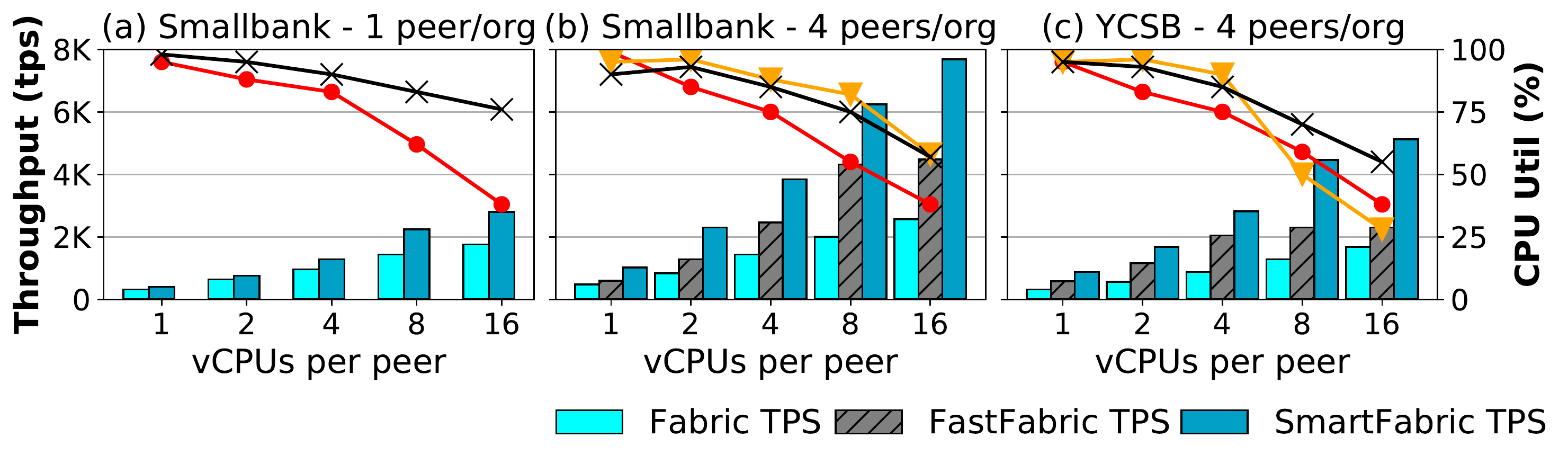}
		\vspace{-.9cm}
		\caption{Vertical Scaling}
		\label{fig:vertical-scaling}
	\end{minipage}
	\begin{minipage}{0.42\textwidth}
        \raggedleft
		\includegraphics[height=85pt]{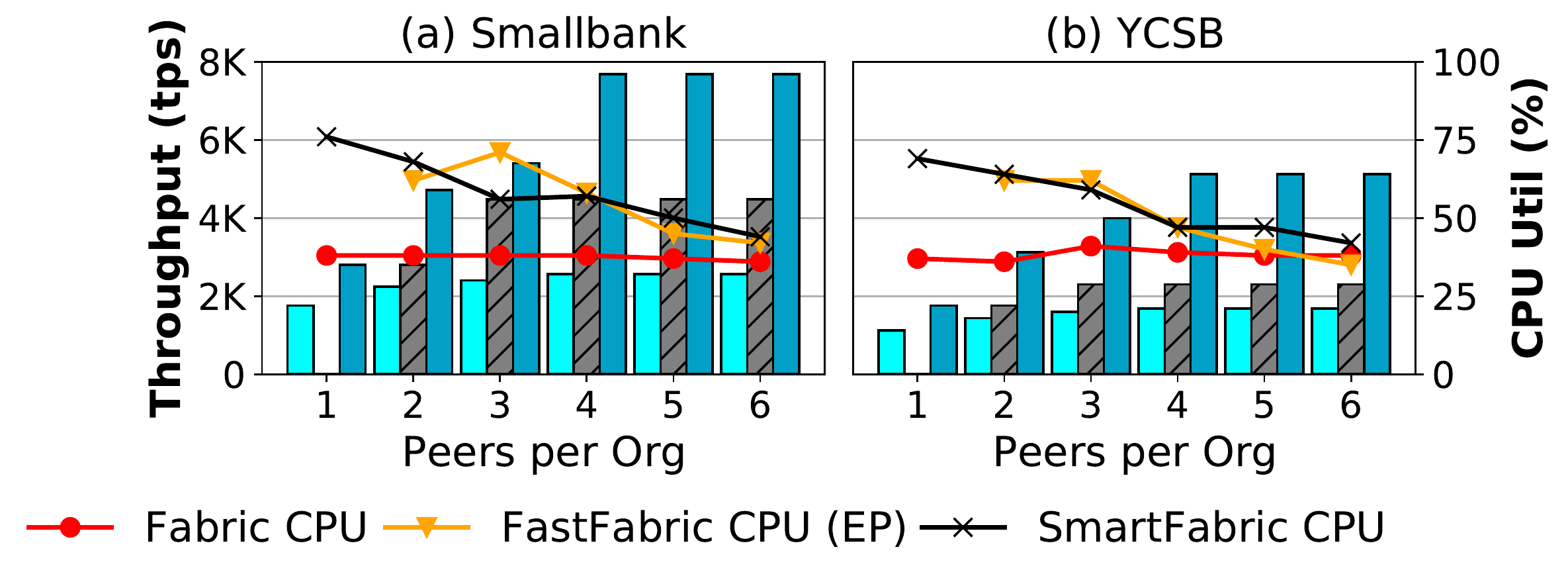}
		\vspace{-.9cm}
		\caption{Horizontal Scaling}
		\label{fig:horizontal-scaling}
	\end{minipage}
	\vspace{-.5cm}
\end{figure*}

\par
Let us assume we have two validation workers, $w_1$ and $w_2$, per peer. Each worker would pick a
transaction with no out-edges, as shown in row one of Table~\ref{table:illus-val-sparse-peer}.
The peer $P_1$ would find $T_1$ and $T_2$ to be valid based on OCC~\cite{occ} check. However, only the
write-set of $T_1$ would be applied to the dirty state while $T_2$ would wait for the validation
result from $P_2$. Similarly, peer $P_3$ would wait for the validation result of $T_3$ from 
$P_1$ and $P_2$. Once all required validation results are shared between peers, $T_2$ would be
marked valid on peer $P_1$, and the dirty state would be updated. However, on peer $P_2$, the
dirty state would not be updated with $T_2$'s write-set as it does not write to $S_2$.
As $T_3$ has a fate dependency on $T_1$, peer $P_1$ would invalidate $T_3$ and share the result
with the other two peers so that they can invalidate it too. Row 6 in Table~\ref{table:illus-val-sparse-peer} presents the final
result.
\vspace{-.4cm}
\subsection{Auto-Scaling Primitives}
\label{scaling}
From \hyperlink{takeaway4}{takeaway 4} in \S\ref{hs-mitigation}, we know that dynamically scaling a Fabric network by adding more peers takes significant time.
A newly added vanilla peer copies, validates, and commits all the blocks sequentially, resulting in a significant time before the State DB is in sync with other peers.
We propose to dramatically reduce the horizontal scale-up time by directly copying a `snapshot' of the State DB at a given block $B$ and then continuing
regular validation and commit of blocks that come after block B.

However, Fabric only stores the latest states in the State DB as of the last committed block, which may get overwritten by future blocks.
For example, Table~\ref{table:db-state} presents the states at state DB after committing each block.
\input{tables/split-peer.tex}
To copy state DB
at block 3, we need to copy the value of $k_1$, $k_2$, and $k_3$ at version $v_1$. As the peer continues to commit blocks, the
latest state changes. After committing block 4, $k_1$ doesn't even exist in state DB, and $k_2$ has a new version $v_2$.
In such cases, the older versions of $k_1$ and $k_2$ can be obtained from write-sets of transactions present in the block store.
To copy states from State DB as of a given block, we add an index of the form \{\textit{smart-contract, block number, transaction number}\} $\mapsto$ a
list of \{\textit{key, isDelete, isDeferred}\}. This tracks state modifications done by every transaction.
As we would be scaling up by adding \textit{sparse peers}, we would only be copying a part of the State DB.
The states to be copied can be identified by doing a range query on this index.
For example, if a newly joined sparse peer wants the state of smart-contract $S_1$ as of block $B$, it can request
the full peer. The full peer runs the range query [\textit{Start}=\{$S_1$, 0, 0\}, \textit{End}=\{$S_1$, $B+1$, 0\}) on the newly added index, and finds the keys modified in this range.
It can then use the states from State DB whose version present in the State DB were created in the requested block range.
If a state $k$ is not in the State DB because it was deleted or modified by a block $B' > B$,
we find the last transaction in the block range $[0, B]$ that wrote $k$. Then, we extract $k$'s value from the transaction's write-set.

Once the State DB is copied, the new peer can start sharing the load. Simultaneously, the admin can update the filter (i.e., removing a contract)
of full peer to make it a sparse peer.
If the new peer will further split into more sparse peers later, we would need to copy the index and block store in the background, which
can happen slowly. Instead, if the new peer needs to be spun up to handle transient load spikes, the State DB copy would suffice.

%% file: tables/dependency-graph.tex
{
		\small{
	 \label{table:rwset}
			\begin{tabular} {| p{.8cm} | p{1.6cm} | p{3.3cm} | p{2cm} |}
        \hline
                    \textbf{Block} & \textbf{Transaction} & \textbf{Read Set} (key, version) & \textbf{Write Set} \\ \hline
                    \multirow{4}{*}{$B_{1}$}  & $T_{1}$ & ($k_{6}, v_{1}$) & ($k_{1}, v_{1} \rightarrow v_{2}$) \\ \cline{2-4}
                     & $T_{2}$ & ($k_{1}, v_{1}$) & ($k_{6}, v_{1} \rightarrow v_{2}$) \\ \cline{2-4}
                     & $T_{3}$ & -                 & ($k_{6}, v_{1} \rightarrow v_{2}$) \\ \cline{2-4}
                     & $T_{4}$ & - & ($k_{5}, v_{1}$) \\ \hline
                    \multirow{2}{*}{$B_{2}$} & $T_{5}$ & range($k_{2}, k_{5}$) & ($k_{7}, v_{1} \rightarrow v_{2}$) \\ \cline{2-4}
                    & $T_{6}$ & ($k_{4}, v_{1}$) & ($k_{4}, v_{1} \rightarrow v_{2}$) \\ \hline
  \end{tabular}
			}
}

%% file: tables/illus-validator.tex
{
	\small{
		\begin{tabular} {| p{.5cm} | p{0.25cm} | p{0.25cm} | p{.25cm} | p{2.3cm} | p {5cm} | p{1.cm} | p{1.cm} | p{3.3cm}|}
			\hline
			\multirow{2}{*}{\hspace{-0.1cm}\textbf{Row}} & \multicolumn{3}{c|}{\textbf{Workers}} &  \multirow{2}{*}{\textbf{Dirty State}} & \multirow{2}{*}{\textbf{State DB}} & \multicolumn{2}{c|}{\textbf{Results}} & \multirow{2}{*}{\textbf{Dependency Graph}} \\ \cline{2-4} \cline{7-8}
			& \textbf{$w_{1}$} & \textbf{$w_{2}$} & \textbf{$w_{3}$} & & & \textbf{Valid} & \textbf{Invalid} & \\ \hline 
			1 & $T_{1}$ & $T_{4}$ & - & - & ($k_{1}$, $v_{1}$), ($k_{4}$, $v_{1}$), ($k_{6}$, $v_{1}$) & - & - & Same as above \\ \hline
			2 & $T_{3}$ & $T_{6}$ & - & ($k_{1}$, $v_{2}$), ($k_{5}$, $v_{1}$) & ($k_{1}$, $v_{1}$), ($k_{4}$, $v_{1}$), ($k_{6}$, $v_{1}$)  & $T_{1}$, $T_{4}$ & $T_{2}$, $T_{5}$ & Only $T_{3}$ \& $T_{6}$, no edges \\ \hline
			3 & - & - & - & ($k_{1}$, $v_{2}$), ($k_{5}$, $v_{1}$)\newline ($k_{6}$, $v_{2}$), ($k_{4}$, $v_{2}$)  & ($k_{1}$, $v_{1}$), ($k_{4}$, $v_{1}$), ($k_{6}$, $v_{1}$) & $T_{1}$, $T_{4}$\newline $T_{3}$, $T_{6}$  & $T_{2}$, $T_{5}$ & Empty \\ \hline
			\multicolumn{9}{|c|}{\hspace{-1cm}\cellcolor{blue!25}Committer commits the block $B_{1}$} \\ \hline
			4 & - & - & - & ($k_4$, $v_2$) & ($k_1$, $v_2$), ($k_4$, $v_1$) ($k_5$, $v_1$), ($k_6$, $v_2$) & $T_6$ & $T_5$ & - \\ \hline
			\multicolumn{9}{|c|}{\hspace{-1cm}\cellcolor{blue!25}Committer commits the block $B_{2}$} \\ \hline
			5 & - & - & - & - & ($k_1$, $v_2$), ($k_4$, $v_2$) ($k_5$, $v_1$), ($k_6$, $v_2$) & - & - & - \\ \hline
		\end{tabular}
	}
}

%% file: tables/transactions-sparse-peer.tex
\definecolor{navyblue}{rgb}{0.0, 0.0, 0.5}
\definecolor{ao(english)}{rgb}{0.0, 0.5, 0.0}
\definecolor{coolblack}{rgb}{0.0, 0.18, 0.39}
\definecolor{darkmagenta}{rgb}{0.55, 0.0, 0.55}
\definecolor{Gray}{gray}{0.85}

{
		\small{
	 \label{table:rwset-sparse-peer}
			\begin{tabular} {| p{.5cm} | p{1.7cm} | p{1.7cm} |}
        \hline
                    \textbf{Tx} & \textbf{Read Set} & \textbf{Write Set} \\ \hline
                    $T_1$ & \textcolor{red}{$S_1 \{(k_7, v_1)\}$} & \textcolor{red}{$S_1 \{(k_7, v_2)\}$} \\ \hline
                    $T_2$ & \textcolor{ao(english)}{$S_2$ \{$(k_3, v_1)$\}} & \textcolor{red}{$S_1$ \{$(k_4, v_2)$\}} \\ \hline
                    $T_3$ & \textcolor{red}{$S_1$ \{$(k_4, v_1)$\}} \newline \textcolor{ao(english)}{$S_2$ \{$(k_5, v_1)$\}} \newline \textcolor{darkmagenta}{$S_3$ \{$(k_6, v_1)$\}} & \textcolor{red}{$S_1$ \{$(k_4, v_2)$\}} \newline \textcolor{ao(english)}{$S_2$ \{$(k_3, v_2)$\}} \newline \textcolor{darkmagenta}{$S_3$ \{$(k_6, v_2)$\}} \\ \hline
  \end{tabular}
			}
}

%% file: tables/statedb-sparse-peer.tex
\definecolor{navyblue}{rgb}{0.0, 0.0, 0.5}
\definecolor{ao(english)}{rgb}{0.0, 0.5, 0.0}
\definecolor{coolblack}{rgb}{0.0, 0.18, 0.39}
\definecolor{darkmagenta}{rgb}{0.55, 0.0, 0.55}
\definecolor{Gray}{gray}{0.85}

{
        \small{
     \label{table:statedb-sparse-peer}
            \begin{tabular} {| p{1.1cm} | p{1.1cm} | p{1.1cm} |}
        \hline
        \textbf{Peer $P_1$} & \textbf{Peer $P_2$} &\textbf{Peer $P_3$} \\ \hline
\textcolor{red}{$S_1$ ($k_4, v_1$) ($k_7, v_1$)} & \textcolor{ao(english)}{$S_2$ ($k_3, v_1$) ($k_5, v_1$)} & \textcolor{darkmagenta}{$S_3$ ($k_6, v_1$)} \\ \hline
  \end{tabular}
            }
}

%% file: tables/illus-sparse-peer.tex
\definecolor{navyblue}{rgb}{0.0, 0.0, 0.5}
\definecolor{ao(english)}{rgb}{0.0, 0.5, 0.0}
\definecolor{coolblack}{rgb}{0.0, 0.18, 0.39}
\definecolor{darkmagenta}{rgb}{0.55, 0.0, 0.55}
\definecolor{Gray}{gray}{0.85}

\newcolumntype{N}{>{\centering}p{0.25cm}}
\newcolumntype{Z}{>{\centering}p{0.2cm}}
\newcolumntype{Y}{>{\centering}p{1.65cm}}
\newcolumntype{X}{>{\centering}p{0.9cm}}
\newcolumntype{M}{>{\centering\arraybackslash}p{2.2cm}}

{

	\small{
		\centering
		\begin{tabular} {|p{.4cm}| Z | Z | Z | Z | Z | Z | X | N | N | N | N | N | N | N | N | Y  | Y | M |}
			\hline
            \multirow{3}{*}{\hspace{-.1cm}Row} & \multicolumn{6}{c|}{\textbf{Workers}} & \multicolumn{3}{c|}{\textbf{Dirty State}} & \multicolumn{3}{c|}{\textbf{Valid Tx}} & \multicolumn{3}{c|}{\textbf{Invalid Tx}} & \multicolumn{3}{c|}{\textbf{Tx waiting for other peers}}  \\ \cline{2-19}
            & \multicolumn{2}{c|}{\textbf{$P_1$}} & \multicolumn{2}{c|}{\textbf{$P_2$}} & \multicolumn{2}{c|}{\textbf{$P_3$}} & \multirow{2}{*}{\textbf{$P_1$}} & \multirow{2}{*}{\textbf{$P_2$}} & \multirow{2}{*}{\textbf{$P_3$}} & \multirow{2}{*}{\textbf{$P_1$}} & \multirow{2}{*}{\textbf{$P_2$}} & \multirow{2}{*}{\textbf{$P_3$}} & \multirow{2}{*}{\textbf{$P_1$}} & \multirow{2}{*}{\textbf{$P_2$}} & \multirow{2}{*}{\textbf{$P_3$}} & \multirow{2}{*}{\textbf{$P_1$}} & \multirow{2}{*}{\textbf{$P_2$}} & \multirow{2}{*}{\textbf{$P_3$}} \\ \cline{2-7} 
            & \textbf{$w_1$} & \textbf{$w_2$} & \textbf{$w_1$} & \textbf{$w_2$} & \textbf{$w_1$} & \textbf{$w_2$} & & & & & & & & & & & & \\ \hline
            1 & $T_1$ & $T_2$ & $T_2$ & - & $T_3$ & - & & & & - & - & - & - & - & - & - & - & - \\ \hline
            2 & -&-&-& - &-& - & \textcolor{red}{($k_7, v_2$)} & & & $T_1$ & - & - & - & - & - & $T_2$ awaits $P_2$ & $T_2$ awaits $P_1$ & $T_3$ awaits $P_1$, $P_2$ \\ \hline
            \multicolumn{8}{|l}{\cellcolor{Gray}\color{blue}$P_1$ informs $P_2$ that $T_2$ is \textcolor{red}{$S_1$}-valid} & \multicolumn{8}{c}{\cellcolor{Gray}\color{blue}$P_2$ informs $P_1$ that $T_2$ is \textcolor{ao(english)}{$S_2$}-valid} & \multicolumn{3}{r|}{\cellcolor{Gray}\color{blue}$P_3$ informs both $P_1$ and $P_2$ that $T_3$ is \textcolor{darkmagenta}{$S_3$}-valid} \\ \hline

            3 & -&-&$T_3$& - &-& - & \textcolor{red}{($k_7, v_2$) ($k_4, v_2$)} & & & $T_1$ $T_2$ & $T_2$ & - & $T_3$ & - & - & - & - & $T_3$ awaits $P_1$, $P_2$ \\ \hline
            \multicolumn{19}{|c|}{\cellcolor{Gray}\color{blue}$P_1$ informs $P_2$ and $P_3$ that $T_3$ is invalid} \\ \hline
            4 & -&-& - & - &-& - & \textcolor{red}{($k_7, v_2$) ($k_4, v_2$)} & & & $T_1$ $T_2$ & $T_2$ & - & $T_3$ & $T_3$ & $T_3$ & - & - & - \\ \hline
		\end{tabular}
	}
}

%% file: tables/split-peer.tex
\begin{wraptable}{l}{0.26\textwidth}
	\vspace{-.3cm}
		\small{
	\centering
			\caption{\small{{KV pairs in State DB}}}
            \vspace{-.35cm}
	 \label{table:db-state}
			\begin{tabular} {| p{1cm} | p{3cm} |} \hline
                \textbf{Block\#} &  \textbf{State DB} \\ \hline
                    1 & $(k_{1}, v_{1})$ \\ \hline
                    2 & $(k_{1}, v_{1}), (k_{2}, v_{1})$ \\ \hline
                    3 & $(k_{1}, v_{1}), (k_{2}, v_{1}), (k_{3}, v_{1})$ \\ \hline
                    4 & $(k_{2}, v_{2}), (k_{3}, v_{1})$ \\ \hline
            \end{tabular}
			}
	\vspace{-.3cm}
\end{wraptable}


%% file: evaluation.tex
\newcommand{\ourApproach}{SmartFabric} 
\def\lc{\left\lceil}   
\def\rc{\right\rceil}

\section{Evaluation}
\label{evaluation}
We implemented our proposals by adding 15K lines of GoLang code to Hyperledger Fabric v1.4.
Hereafter, we refer to our approach as \textbf{\ourApproach}.
We compare {\ourApproach} with vanilla Fabric and FastFabric~\cite{fastfabric}, which claims to achieve 20K tps.
A follow-up work~\cite{xox} showed that the stable implementation~\cite{fastfabric-github} of FastFabric achieved 14K tps on a benchmark similar to Smallbank.
Briefly, in order to scale, FastFabric separates a peer's roles to 3 separate node types: Endorser Peers (EP),
FastPeer (FP) and Storage Peers (SP). Only SP commits both blocks and state on the disk. Other nodes only store state in RAM.
Note that we have applied one of their optimizations (block deserialization cache) to both vanilla Fabric (as mentioned in \S\ref{motivation}) and our implementation.
First, we focus our evaluation on the following guiding questions:
\begin{enumerate}[label=\textbf{Q\arabic*.},itemsep=-0pt, topsep=-1pt]
	\item Given a fixed cost of infrastructure, how much higher throughput can {\ourApproach} provide?
	\item Given a required throughput, by how much does {\ourApproach} reduce the infrastructure cost?
\end{enumerate}
Next, we inspect the internals to show how each of the proposed optimizations individually performs.
For experiments, we use the same default configurations described in \S\ref{motivation}.



\subsection{Vertical Scaling}
To study the impact of vertical scaling on SmartFabric and FastFabric, we considered the following two scenarios: (1) single
peer per organization; (1) four peers per organization.
In scenario (1), \textit{sparse peers} and FastFabric are not applicable as they require at least 2 peers per organization.

\textbf{(1) Single peer per organization}. 
Figure~\ref{fig:vertical-scaling}(a) plots the impact of number of vCPUs on the throughput and CPU utilization under the Smallbank workload.
{\ourApproach} provided 1.4$\times$ higher throughput (on average) than Fabric for the same infrastructure cost.
Further, with an increase in the number of vCPUs from 1 to 16, {\ourApproach}'s throughput improved 7$\times$ compared to 5.5$\times$ improvement with Fabric.
Similarly, {\ourApproach} required only 8 vCPUs to provide higher than Fabric's throughput with 16 vCPUs, i.e., better performance at half the cost.
We obtained similar results with YCSB but omitted the plot for brevity.
This improvement is because Fabric underutilizes the CPU beyond a point while {\ourApproach} maintains a high utilization.
For example, with 16 vCPUs, {\ourApproach} demonstrates 76\% CPU utilization versus 36\% utilization with Fabric.

\textbf{(2) Four peers per organization}.
Figures~\ref{fig:vertical-scaling}(b) and \ref{fig:vertical-scaling}(c) plot the impact of number of vCPUs on the throughput and CPU utilization
for both smallbank and YCSB, respectively.
{\ourApproach} provided 2.7$\times$ and 1.7$\times$ higher throughput (on average) than Fabric and FastFabric, respectively, for the same infrastructure cost.
For example, with 16 vCPUs, {\ourApproach} provided 7680 tps versus Fabric's 2560 tps and FastFabric's 4480 tps.
Further, Fabric's peak throughput with 16 vCPU was met by FastFabric with 4 vCPUs, while {\ourApproach} provided comparable performance with just 2 vCPUs (87\% lower cost).
{\ourApproach} showed these improvements due to a combination of pipelined executions and sparse peers.
There are two interesting observations with FastFabric's CPU utilization:
(1) the CPU utilization of FastFabric and SmartFabric is almost the same under Smallbank, yet there is a performance gap;
(2) the CPU utilization of FastFabric suddenly dropped when the number of vCPUs > 4 under YCSB.
The reasons for these interesting behaviors are explained in the next section.

\subsection{Horizontal Scaling comparison}
\begin{figure*}[t]
	\begin{minipage}{0.49\textwidth}
		\centering
		\includegraphics[height=65pt]{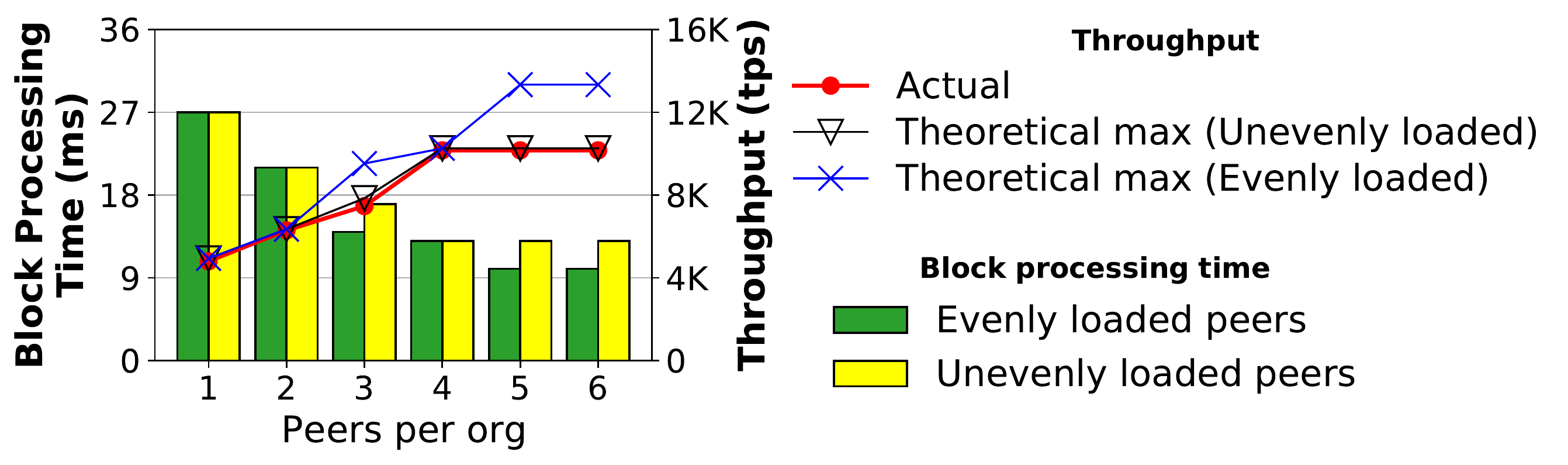}
	\vspace{-.5cm}
        \caption{\small{Effect of imbalanced load on Sparse Peers}}
		\label{fig:sp-block-proc-time}
	\end{minipage}
	\begin{minipage}{0.49\textwidth}
		\raggedleft
		\includegraphics[height=65pt]{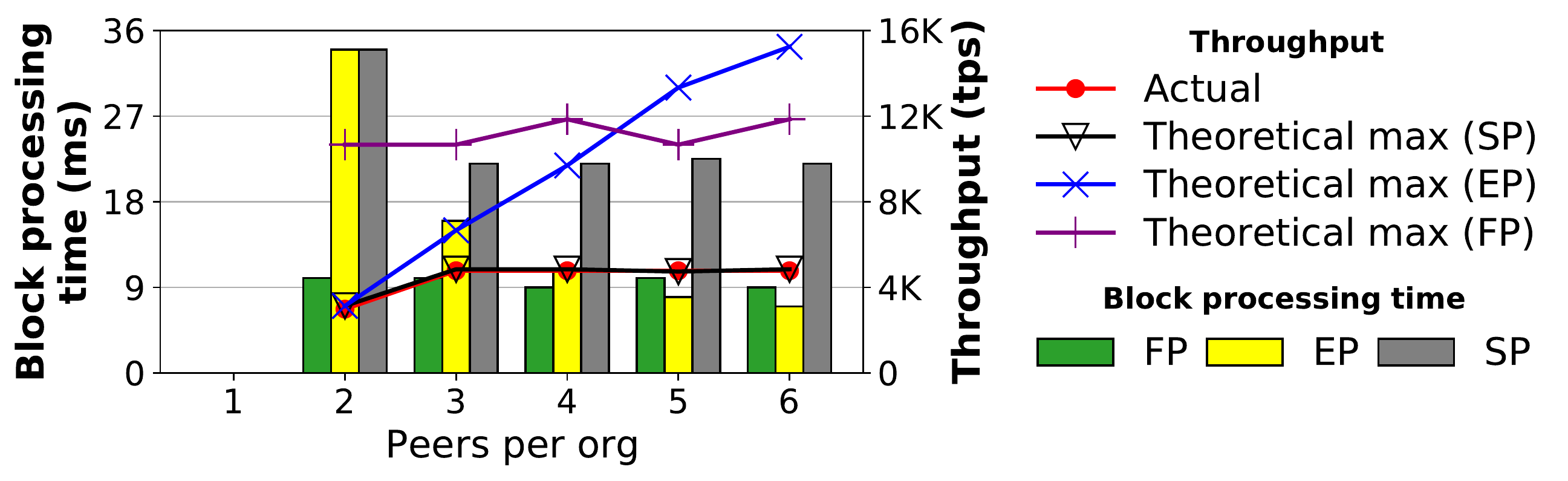}
	\vspace{-.5cm}
        \caption{\small{FastFabric's block processing time}}
		\label{fig:ff-block-proc-time}
	\end{minipage}
	\vspace{-0.3cm}
\end{figure*}
\begin{figure*}[t]
	\begin{minipage}{0.3\textwidth}
		\raggedright
		\includegraphics[height=85pt]{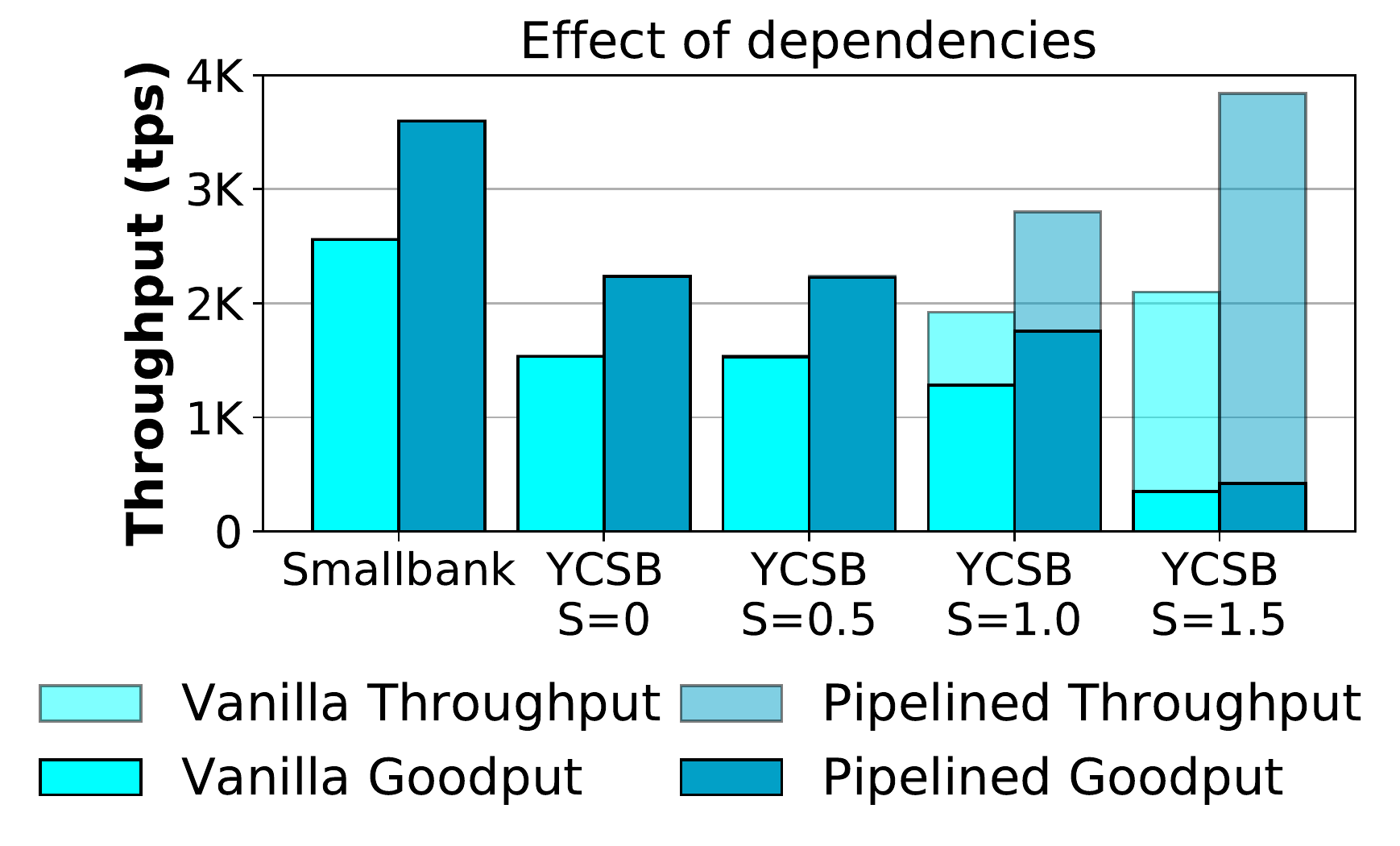}
		\vspace{-.5cm}
		\caption{{Effect of dependencies}}
  		\label{fig:dep-perf}
	\vspace{-0.3cm}
	\end{minipage}
	\hspace{0.1cm}
	\begin{minipage}{0.32\textwidth}
		\centering
		\includegraphics[height=75pt]{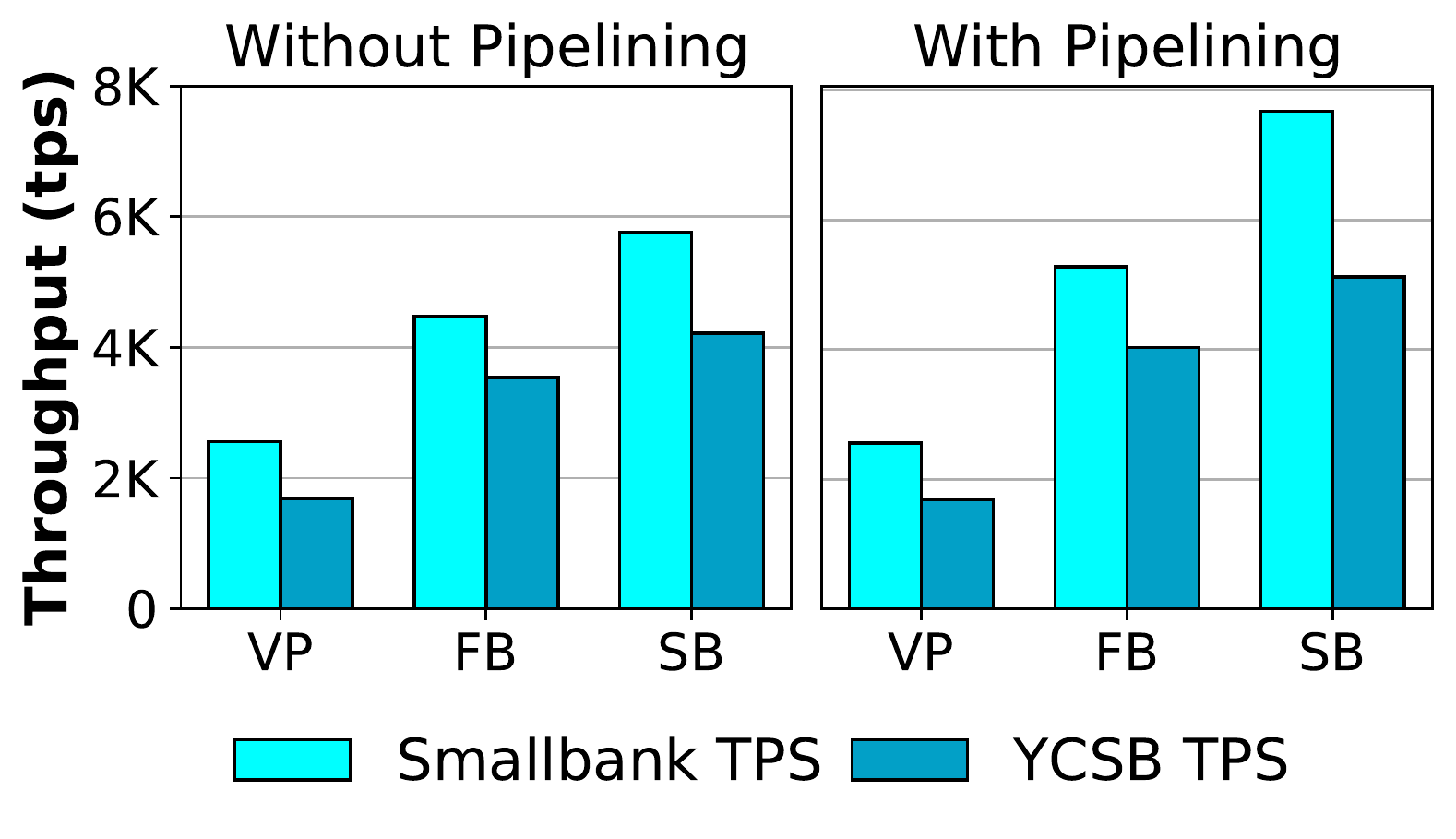}
		\vspace{-.5cm}
		\caption{\small{Full Blocks (FB) \& Sparse Blocks (SB) vs Vanilla Peer (VP)}}
		\label{fig:full-vs-sparse-blocks}
	\vspace{-0.3cm}
	\end{minipage}
	\hspace{0.05cm}
	\begin{minipage}{0.35\textwidth}
		\raggedright
		\includegraphics[height=80pt]{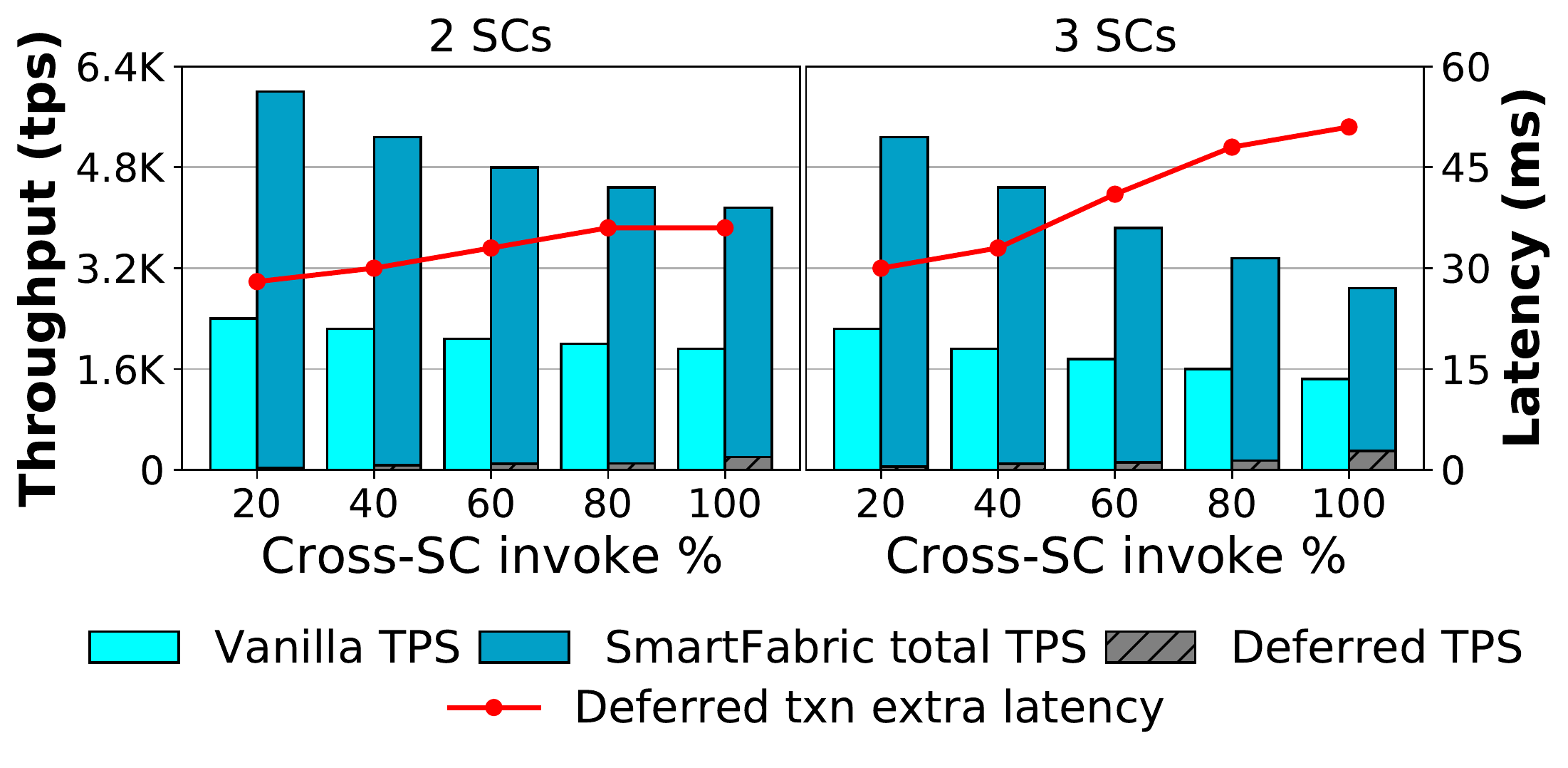}
		\vspace{-.5cm}
		\caption{\small{Distributed transactions performance}}
		\label{fig:cc2cc}
	\vspace{-0.3cm}
	\end{minipage}
\end{figure*}

Figure~\ref{fig:horizontal-scaling} plots the impact of number of peers in an organization on the throughput and CPU utilization.
Each peer had 16 vCPUs. For FastFabric, when there were $n$ nodes, we configured them as 1 FP, 1 SP, and $(n-2)$ EP nodes.
FastFabric requires a single dedicated FP node. We kept only one SP because, like vanilla Fabric, each SP node would do the same work
of committing the entire block to disk. Adding more EP nodes helps by allowing higher transaction endorsement.
When there were only 2 nodes, we overlapped EP and SP roles on the same node because FP must be a dedicated node.
For {\ourApproach}, we evenly divided $m$ smart contracts over $n$ nodes.
To every node's \textit{filter}, we first added $\left \lfloor\frac{m}{n}\right \rfloor$ contracts.
We then added one \textit{additional} contract to $m \mod{n}$ nodes. We call these nodes `unevenly' loaded, while others are referred to as `evenly loaded'.
In our case, $m = 8$. When $n=6$, each node gets 1 contract, and two nodes get an additional contract.

{\ourApproach} provided 2.5$\times$ and 1.6$\times$ higher throughput (on average) than Fabric and FastFabric, respectively, for the same number of peers per organization.
Further, with just a single node, {\ourApproach} achieved more throughput than Fabric achieved using 4 nodes, i.e., better performance with $75\%$ lower cost.
Similarly, {\ourApproach} provided higher throughput with 2 nodes than FastFabric with 3 or more nodes.
Interestingly, no approach provided higher throughput beyond 4 peers per organization.
As we have already explained Fabric's behavior in Section~\ref{background-and-motivation}, we explain the behavior of FastFabric and {\ourApproach} next.

FastFabric did not benefit by adding even the fourth peer as it got bottlenecked by the SP node.
In Section~\ref{background-and-motivation}, we showed that a vanilla peer that does not endorse transactions
could achieve 3300 tps. An SP node is exactly like this, except it gets pre-validated transactions
and only commits them. Therefore, its peak throughput increased to 4480 tps.
Figure~\ref{fig:ff-block-proc-time} shows the block processing times along with the theoretical maximum throughput that each node can provided. 
The actual throughput matched the theoretical maximum of SP nodes.

{\ourApproach} also did not show improvement on adding more than 4 peers, but the reason is different.
With four nodes, each node committed transactions for two smart-contracts, and the IO-load was balanced. 
With six nodes, four nodes were evenly loaded because they managed a single contract. The other two were unevenly loaded as
they managed two contracts. Therefore, in this case, the slowest node was still similar to the case with four nodes.
This can be seen in Figure~\ref{fig:sp-block-proc-time}. The theoretical max of evenly
loaded peers increased as the number of contracts they managed decreased. Nevertheless, the actual
throughput was dictated by unevenly loaded peers.
Thus, in reality, the benefit of the \textit{sparse peer} optimization would depend upon the number of
contracts, their load distribution characteristics, and the number of peers in the organization.
%
%


\subsection{Inspecting internals}
We now perform an ablation study by looking at individual optimizations. All experiments were done with
4 peers per organization (16 peers in total), each having 16 vCPUs.

\textbf{(1) Pipelined Validation and Commit Phases}.
In this experiment, all peers were full peers.
We first compare the peak performances achieved on the Smallbank workload.
Figure~\ref{fig:dep-perf} plots the throughput achieved with the pipelined execution against vanilla Fabric. 
As expected, there was a 1.4$\times$ improvement over vanilla Fabric while increasing the CPU utilization from 40\% to 60\%.
The validation manager was so efficient that the committer never got blocked.
The size of the \textit{result-map} was always larger than 200. This is because the
time taken by the committer ($\approx$22 ms at 3600 endorsement requests per second---eps)
was always higher than the time taken by validators. Further, the end-to-end commit latency
(validation + commit) for a block was reduced from 39 \textit{ms}
(at 2560 eps) to 27 ms (at 3600 eps).
\par
To study the effect of different degrees of dependencies between transactions, in YCSB, we varied the
skewness of the Zipf distribution, which was used to
select the keys: we went from an s-value=0.0 (uniform) to an s-value=1.5 (highly skewed) in
steps of 0.5. Figure~\ref{fig:dep-perf} plots the throughput and goodput achieved with pipelined execution
against vanilla Fabric. The reason for considering goodput is that certain transactions get
%
%
\input{tables/mvsg.tex}

invalidated due to serializability conflicts. As expected, the pipelined execution outperformed
vanilla Fabric when the s-value was low. As the s-value increased, more transactions were
invalidated, reducing the goodput and performance gain with pipelining. Further, it is
interesting to note that the throughput increased with a decrease in the goodput as the peer
did not spend much time on the commit.
		

\textbf{(2) Sparse Peer with Full and Sparse Blocks}.
We evaluate the performance of two variants of sparse peer proposed in \S\ref{sparse-peer}.
Each organization hosted 4 sparse peers where the filter of each sparse peer contained
only 2 non-overlapping smart-contracts. Figure~\ref{fig:full-vs-sparse-blocks}(a) plots the throughput
achieved with both variants of sparse peers against a network where each organization
hosted 4 vanilla peers. As expected, the throughput increased significantly by 2.4$\times$
with sparse peers for both workloads.
Compared to the sparse
peer processing full blocks, the sparse peer processing sparse block achieved higher throughput
due to the reduced IO operation.
\label{eval:sparse}
Figure~\ref{fig:full-vs-sparse-blocks}(b) plots the throughput achieved
with the combination of sparse peer and pipelined execution against vanilla Fabric. The throughput increased
significantly to 7680 tps, i.e., 3$\times$ that of vanilla Fabric.

\textbf{(3) Distributed Simulation and Validation}. When transactions invoke multiple smart-contracts,
{\ourApproach} employs distributed simulation and validation. {To study
our proposed system's performance} in the presence of distributed transactions, we submitted transactions that invoked multiple
smallbank contracts. Figure~\ref{fig:cc2cc} compares the throughput of vanilla Fabric
against {\ourApproach} under a varying mix of cross-contract invocation transactions.
We consider two scenarios where cross-contract transactions invoke 2 and 3 contracts, respectively.
As expected, {\ourApproach}'s performance degraded as the percentage of transactions invoking multiple contracts increased.
This is because, with more transactions invoking multiple contracts, the amount of work to be done at each peer increased.
However, even with 100\% cross-contract transactions, {\ourApproach} outperformed vanilla Fabric by 2$\times$ in both scenarios.

Interestingly, the number of transactions deferred per second was not substantial either.
Even when 100\% of transactions invoked two and three contracts, only $\approx 200$ tps and $\approx 400$ tps respectively got deferred.
In other words, merely 5\% to 10\% of issued transactions were getting deferred even when every transaction
required distributed validation.
The low deferral rate is because, with pipelined validation and commit phases,
transactions were getting validated much earlier than their results were required by the committer.
The extra latency experienced by deferred transactions ranged from 30 to 50 ms. Given that blockchain transactions
already take several hundreds of milliseconds~\cite{hf}, this is quite a small penalty to pay
for a big improvement in throughput. However, note that this duration is in the order of one or two block processing
times in {\ourApproach}. Without deferring, the average block processing duration would increase 2$\times$ or worse due to waiting,
thus drastically reducing the peak throughput. That would be a significant penalty to pay for not deferring 5\% to 10\% of transactions.
Thus, deferring transactions is key to this performance.

\begin{wrapfigure}{l}{0.26\textwidth}
	\centering
	\vspace{-0.42cm}
	\includegraphics[width=0.3\textwidth]{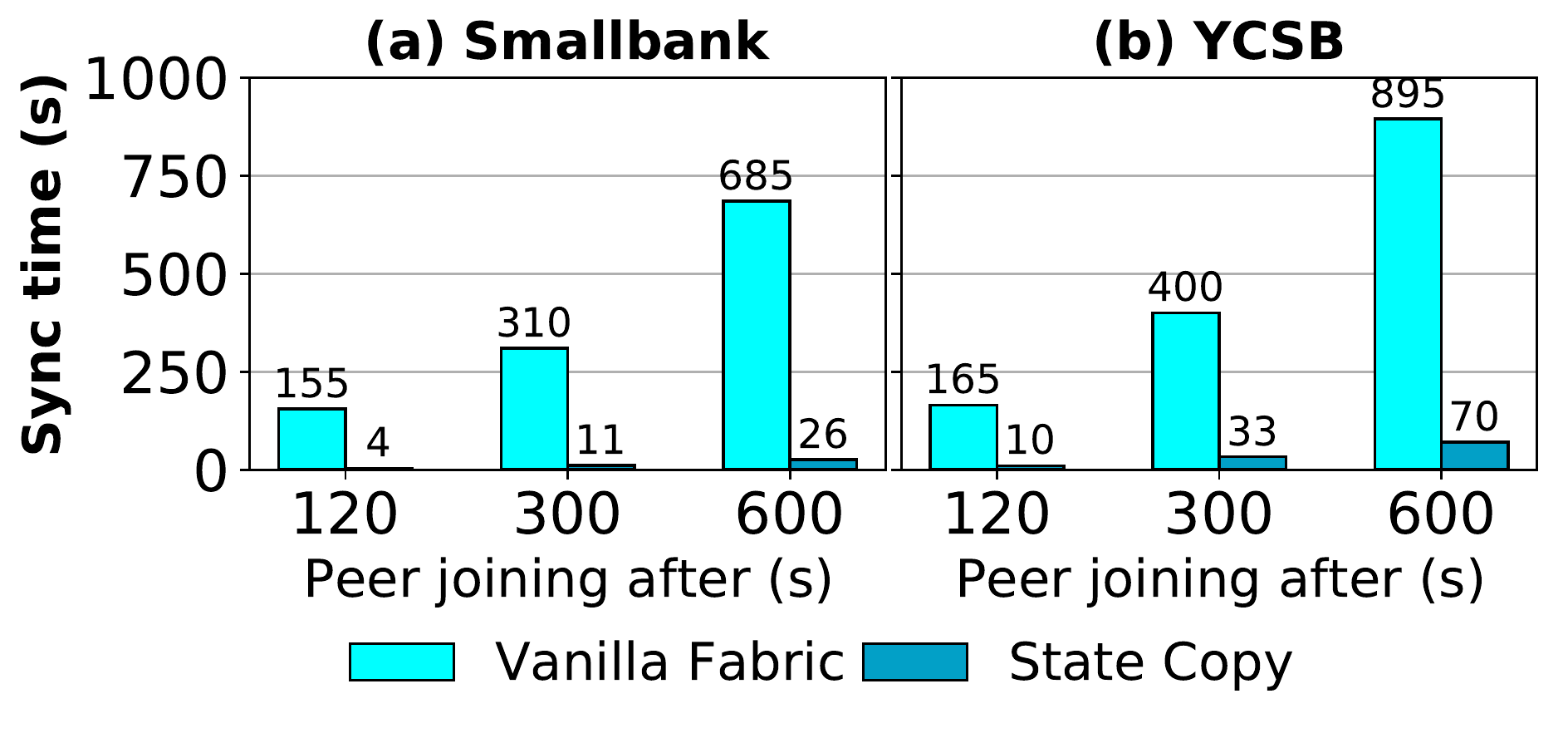}
	\vspace{-.9cm}
	\caption{\small{New peer sync time}}
	\label{fig:adding-peer}
	\vspace{-0.45cm}
\end{wrapfigure}

\textbf{(4) Scaling Up and Down}.
\label{eval:autoscaling}
We evaluate the dynamic scaling appro{a}ch discussed in \S\ref{scaling}. Figure~\ref{fig:adding-peer}
plots the time taken for a new peer to copy states from {another}. To perform a fair
comparison with vanilla, we added a new sparse peer with all
smart-contracts (the equ{i}valent of a full peer) and generated the same load on existing peers.
{\ourApproach}{ provided a multifold reduction than a vanilla peer's sync time at various minutes}.
This is because {\ourApproach} copied a much smaller
amount of data. 
In general, the size of block store is several times higher than the size of state DB.
As we directly copy required states from the state DB, the time taken to add a new peer reduced significantly.
In vanilla Fabric, we could turn off a peer for scaling down a network. Here, we have to pay a
small penalty. Naturally, scaling down sparse peers would require copying states from the sparse peers
to other peers. The time to copy the states was similar to the scale-up time, which is quite small.
Compared to the benefits of sparse peers, this penalty is insignificant.

\par

%% file: tables/mvsg.tex
{
	\begin{table*}[t]
	\vspace{-.2cm}
		\small{
	\centering
	\caption{\small{Comparison of our dependency graph with dependency graphs used in prior arts in permissioned blockchain}}
	\vspace{-.3cm}
	 \label{table:mvsg}
        \begin{tabular} {| p{2.5cm} | p{2.9cm} | p{1.8cm} | p{2.5cm} | p{3.4cm} | p{2.3cm} |}
        \hline
            & \textbf{Our dependency graph} &\textbf{Par\cite{parblockchain}}  & \textbf{Fabric++}~\cite{perf-ssi} & \textbf{FabricSharp}~\cite{tps} & \textbf{XOX-Fabric}~\cite{xox} \\ \hline
            \textit{goal} & To validate transactions in parallel, and respecting the predetermined commit order & To execute transactions parallely & To reorder transactions to reduce the transaction abort rate within a block & To abort unserializable transactions before ordering and to avoid transaction aborts within a block & To re-execute conflicting transactions to reduce abort rate \\ \hline
            \textit{blockchain flavor} & EOV & OE & \multicolumn{3}{c|}{EOV} \\ \hline
            \textit{dependencies tracked} & rw, wr, ww, pr, ep-rw, ep-wr, ep-ww & rw, wr, ww & rw & \multicolumn{2}{c|}{rw, wr, ww} \\ \hline
            \textit{dependency} & across blocks & \multicolumn{2}{c|}{single block} & \multicolumn{2}{c|}{across blocks} \\ \hline
            \textit{cycles} & impossible & \multicolumn{3}{c|}{possible} & impossible \\ \hline
            \textit{supported queries} & get, put, range &  \multicolumn{4}{c|}{get, put} \\ \hline
  \end{tabular}
	\vspace{-.5cm}
}
\end{table*}
}

%% file: relatedwork.tex
\section{Related Work}
\label{relatedwork}
In this section, we cover the existing work that improved the performance of
Hyperledger Fabric. None of them have studied the performance using
different scaling techniques. \emph{While many past approaches (~\cite{parblockchain}, ~\cite{perf-ssi}, ~\cite{tps}, ~\cite{xox})
track dependencies between transactions, they use them differently.
Table~\ref{table:mvsg} present a detailed comparison of our dependency graph with dependency graphs used in prior arts in permissioned blockchain.}

\par
Thakkar \textit{et al.} ~\cite{perf-mascots} conducted a comprehensive performance study
and found bottlenecks in Fabric v1.0 and provided guidelines to design applications
and operate the network to attain a higher throughput. Further{,} they implemented a few
optimizations on the peer process. These optimizations have already been included in
Fabric v1.4, and hence our work builds upon this work.
\par

Sharma \textit{et al.}~\cite{perf-ssi} used ideas from the database literature to reorder transactions within a block
by analzing their dependencies, with the goal of reducing transaction aborts.
Ruan \textit{et al.}~\cite{tps} extended ~\cite{perf-ssi} to abort unserializable transactions before ordering
and to altogether avoid transaction aborts within a block by reordering them.
These techniques are orthogonal to our work as
we do not modify the ordering of transactions. We focus on pipelined
execution of different phases and to avoid redundant work within an organization.
\par

Gorenflo \textit{et al.}~\cite{fastfabric} proposed FastFabric that includes various optimizations
such as replacing the state DB with a hash table, storing blocks in a separate server, separating the committer
and endorser into different servers, parallelly validating the transactions headers, and caching the
unmarshed blocks to reach a throughput of 20000 tps. However, we believe that many of these
optimizations are not practical for {a} production environment. For example, a state DB is must to support
range queries and persist all {current} states (which would help to recover a peer quickly after a failure).
The proposed
approach assumes the transactions to have no read-write conflicts. Moreover, their work is orthogonal
to ours as they do not (1) execute the validation and commit phases in a pipelined manner;
(2) parallelly validate transaction during the serializability check;
(3) have a concept of sparseness in a peer; (4) provide a framework
to scale up a Fabric network quickly. Note, we have adopted the block cache proposed
in this work{,} as mentioned in \S~\ref{motivation}.
\par
Gorenflo \textit{et al.}~\cite{xox} extended FastFabric~\cite{fastfabric} to introduce post-order
execution of transactions to reduce the transaction aborts. They
constructed a dependency graph at the peer with rw, ww, wr dependencies. Whenever there was a
conflict between a transaction, they re-exeucted the patch-up code passed with the transaction
to reduce the transaction aborts. This work is orthogonal to our approach.


Ruan \textit{et al.}~\cite{db-vs-bc} compared the performance of two permissioned blockchain
platforms---Hyperledger Fabric \& Quorum with distributed databases, namely TiDB,
and etcd. However, they did not study the impact of vertical and horizontal scaling on Hyperledger
Fabric.

%% file: ms.bbl
\begin{thebibliography}{10}

\bibitem{alipay}
Alipay press release.
  https://www.alibabagroup.com/en/news/press\_pdf/p180201.pdf.

\bibitem{aws-cost}
Amazon aws cost estimator. https://calculator.aws/.

\bibitem{azure-cost}
Azure cost estimator. ihttps://azure.microsoft.com/en-in/pricing/calculator/.

\bibitem{quorum}
Consensus quorum. https://consensys.net/quorum/.

\bibitem{electrumdark}
Electrum decentralized marketplace. https://electrumdark.co/.

\bibitem{forbes}
Forbes blockchain 50 2021.
  https://www.forbes.com/sites/michaeldelcastillo/2021/02/02/blockchain-50/?sh=113b8b9231cb.

\bibitem{google-cost}
Google cloud cost estimator. https://cloud.ibm.com/estimator/review.

\bibitem{grpc}
grpc. https://grpc.io/.

\bibitem{ibm-cost}
Ibm cloud cost estimator. https://cloud.google.com/products/calculator.

\bibitem{open-bazaar}
Open bazaar decentralized marketplace. https://openbazaar.org/.

\bibitem{origami-network}
Origami network decentralized marketplace. https://ori.network/.

\bibitem{particl}
particl decentralized marketplace. https://particl.io/.

\bibitem{token2}
Understanding digital tokens: Market overviews and proposed guidelines for
  policymakers and practitioners by token alliance, chamber of digital
  commerce.
  https://morningconsult.com/wp-content/uploads/2018/07/token-alliance-whitepaper-web-final.pdf.

\bibitem{visa-annual-report}
Visa annual report.
  https://s1.q4cdn.com/050606653/files/doc\\\_financials/annual/2018/visa-2018-annual-report-final.pdf.

\bibitem{smallbank}
M.~{Alomari}, M.~{Cahill}, A.~{Fekete}, and U.~{Rohm}.
\newblock The cost of serializability on platforms that use snapshot isolation.
\newblock In {\em 2008 IEEE 24th International Conference on Data Engineering},
  pages 576--585, April 2008.

\bibitem{parblockchain}
M.~J. {Amiri}, D.~{Agrawal}, and A.~E. {Abbadi}.
\newblock Parblockchain: Leveraging transaction parallelism in permissioned
  blockchain systems.
\newblock 2019.

\bibitem{hf}
E.~Androulaki, A.~Barger, V.~Bortnikov, C.~Cachin, K.~Christidis, A.~De~Caro,
  D.~Enyeart, C.~Ferris, G.~Laventman, Y.~Manevich, S.~Muralidharan, C.~Murthy,
  B.~Nguyen, M.~Sethi, G.~Singh, K.~Smith, A.~Sorniotti, C.~Stathakopoulou,
  M.~Vukoli\'{c}, S.~W. Cocco, and J.~Yellick.
\newblock Hyperledger fabric: A distributed operating system for permissioned
  blockchains.
\newblock In {\em Proceedings of the Thirteenth EuroSys Conference}, EuroSys
  '18, pages 30:1--30:15, New York, NY, USA, 2018. ACM.

\bibitem{ansi-isolation}
H.~Berenson, P.~Bernstein, J.~Gray, J.~Melton, E.~O'Neil, and P.~O'Neil.
\newblock A critique of ansi sql isolation levels.
\newblock In {\em Proceedings of the 1995 ACM SIGMOD International Conference
  on Management of Data}, SIGMOD '95, page 1–10, New York, NY, USA, 1995.
  Association for Computing Machinery.

\bibitem{ethereum}
V.~Buterin.
\newblock Ethereum: A next-generation smart contract and decentralized
  application platform, 2014.
\newblock Accessed: July 31, 2019.

\bibitem{token1}
Y.~Chen.
\newblock Blockchain tokens and the potential democratization of
  entrepreneurship and innovation.
\newblock {\em Business Horizons}, 61(4):567 -- 575, 2018.

\bibitem{ycsb}
B.~F. Cooper, A.~Silberstein, E.~Tam, R.~Ramakrishnan, and R.~Sears.
\newblock Benchmarking cloud serving systems with ycsb.
\newblock In {\em Proceedings of the 1st ACM Symposium on Cloud Computing},
  SoCC10, pages 143--154, New York, NY, USA, 2010. Association for Computing
  Machinery.

\bibitem{perf-sharding}
H.~Dang, T.~T.~A. Dinh, D.~Loghin, E.-C. Chang, Q.~Lin, and B.~C. Ooi.
\newblock Towards scaling blockchain systems via sharding.
\newblock In {\em Proceedings of the 2019 International Conference on
  Management of Data}, SIGMOD '19, pages 123--140, New York, NY, USA, 2019.
  ACM.

\bibitem{perf-priority}
S.~Goel, A.~Singh, R.~Garg, M.~Verma, and P.~Jayachandran.
\newblock Resource fairness and prioritization of transactions in permissioned
  blockchain systems (industry track).
\newblock In {\em Proceedings of the 19th International Middleware Conference
  Industry}, Middleware '18, pages 46--53, New York, NY, USA, 2018. ACM.

\bibitem{fastfabric-github}
C.~Gorenflo.
\newblock Fastfabric 1.4 implementation:
  https://github.com/cgorenflo/fabric/tree/fastfabric-1.4.

\bibitem{xox}
C.~{Gorenflo}, L.~{Golab}, and S.~{Keshav}.
\newblock Xox fabric: A hybrid approach to blockchain transaction execution.
\newblock In {\em 2020 IEEE International Conference on Blockchain and
  Cryptocurrency (ICBC)}, pages 1--9, 2020.

\bibitem{fastfabric}
C.~{Gorenflo}, S.~{Lee}, L.~{Golab}, and S.~{Keshav}.
\newblock Fastfabric: Scaling hyperledger fabric to 20,000 transactions per
  second.
\newblock In {\em 2019 IEEE International Conference on Blockchain and
  Cryptocurrency (ICBC)}, pages 455--463, May 2019.

\bibitem{perf-noblock}
Z.~Istv\'{a}n, A.~Sorniotti, and M.~Vukoli\'{c}.
\newblock Streamchain: Do blockchains need blocks?
\newblock In {\em Proceedings of the 2Nd Workshop on Scalable and Resilient
  Infrastructures for Distributed Ledgers}, SERIAL'18, pages 1--6, New York,
  NY, USA, 2018. ACM.

\bibitem{occ}
H.~T. Kung and J.~T. Robinson.
\newblock On optimistic methods for concurrency control.
\newblock {\em ACM Trans. Database Syst.}, 6(2):213--226, June 1981.

\bibitem{endorsement1}
Y.~{Manevich}, A.~{Barger}, and Y.~{Tock}.
\newblock Endorsement in hyperledger fabric via service discovery.
\newblock {\em IBM Journal of Research and Development}, 63(2/3):2:1--2:9,
  March 2019.

\bibitem{perf-concurrency}
H.~Meir, A.~Barger, and Y.~Manevich.
\newblock Increasing concurrency in hyperledger fabric.
\newblock In {\em Proceedings of the 12th ACM International Conference on
  Systems and Storage}, SYSTOR '19, pages 179--179, New York, NY, USA, 2019.
  ACM.

\bibitem{bitcoin}
S.~Nakamoto.
\newblock Bitcoin: A peer-to-peer electronic cash system, Dec 2008.
\newblock Accessed: July 31, 2019.

\bibitem{raft}
D.~Ongaro and J.~Ousterhout.
\newblock In search of an understandable consensus algorithm.
\newblock In {\em Proceedings of the 2014 USENIX Conference on USENIX Annual
  Technical Conference}, USENIX ATC'14, pages 305--320, Berkeley, CA, USA,
  2014. USENIX Association.

\bibitem{db-vs-bc}
P.~Ruan, T.~T.~A. Dinh, D.~Loghin, M.~Zhang, G.~Chen, Q.~Lin, and B.~C. Ooi.
\newblock Blockchains vs. distributed databases: Dichotomy and fusion.
\newblock In {\em Proceedings of the 2021 ACM SIGMOD International Conference
  on Management of Data}, SIGMOD '21, page 543–557, 2021.

\bibitem{tps}
P.~Ruan, D.~Loghin, Q.-T. Ta, M.~Zhang, G.~Chen, and B.~C. Ooi.
\newblock A transactional perspective on execute-order-validate blockchains.
\newblock In {\em Proceedings of the 2020 ACM SIGMOD International Conference
  on Management of Data}, SIGMOD '20, page 543–557, New York, NY, USA, 2020.
  Association for Computing Machinery.

\bibitem{perf-ssi}
A.~Sharma, F.~M. Schuhknecht, D.~Agrawal, and J.~Dittrich.
\newblock Blurring the lines between blockchains and database systems: The case
  of hyperledger fabric.
\newblock In {\em Proceedings of the 2019 International Conference on
  Management of Data}, SIGMOD '19, pages 105--122, New York, NY, USA, 2019.
  ACM.

\bibitem{perf-mascots}
P.~{Thakkar}, S.~{Nathan}, and B.~{Viswanathan}.
\newblock Performance benchmarking and optimizing hyperledger fabric blockchain
  platform.
\newblock In {\em 2018 IEEE 26th International Symposium on Modeling, Analysis,
  and Simulation of Computer and Telecommunication Systems (MASCOTS)}, pages
  264--276, Sep. 2018.

\end{thebibliography}
